\documentclass[pra,reprint,superscriptaddress]{revtex4-1}

\usepackage{amsmath, amsfonts,braket,graphicx}
\usepackage[colorlinks=true,citecolor=blue,linkcolor=magenta]{hyperref}

\usepackage[normalem]{ulem}
\usepackage[dvipsnames]{xcolor}

\global\long\def\tr{\mathrm{tr}}

\newcommand{\dg} {{\dagger}}
\newcommand{\pd} {{\phantom\dagger}}

\newcommand{\ci}[1] {{c_{#1}^{\pd}}}
\newcommand{\cid}[1] {c_{#1}^\dg}

\newcommand{\ql}{\mathcal{L}}
\newcommand{\qs}{\mathcal{S}}
\newcommand{\qr}{\mathcal{R}}

\newcommand{\qi}{\mathcal{I}}

\newcommand{\qc}{\mathcal{C}}
\newcommand{\qb}{\mathcal{B}}
\newcommand{\qw}{\mathcal{W}}

\newcommand{\zi}{\chi}
\newcommand{\zI}{X}
\newcommand{\zid}{x}

\newcommand\trick[1]{}

\begin{document}

\title{Performance of Reservoir Discretizations in Quantum Transport Simulations}
\author{Justin E. Elenewski} 
\affiliation{Biophysical and Biomedical Measurement Group, Microsystems and Nanotechnology Division, Physical Measurement Laboratory, National Institute of Standards and Technology, Gaithersburg, MD, USA}
\affiliation{Institute for Research in Electronics and Applied Physics, University of Maryland, College Park, MD, USA}
\author{Gabriela W\'{o}jtowicz}
\affiliation{Jagiellonian University, Institute of Theoretical Physics, \L{}ojasiewicza 11, 30-348 Krak\'{o}w, Poland}
\author{Marek M. Rams}
\email{marek.rams@uj.edu.pl}
\affiliation{Jagiellonian University, Institute of Theoretical Physics, \L{}ojasiewicza 11, 30-348 Krak\'{o}w, Poland}
\author{Michael Zwolak}
\email{mpz@nist.gov}
\affiliation{Biophysical and Biomedical Measurement Group, Microsystems and Nanotechnology Division, Physical Measurement Laboratory, National Institute of Standards and Technology, Gaithersburg, MD, USA}

\begin{abstract}
\par 
Quantum transport simulations often use explicit, yet finite, electronic reservoirs.  These should converge to the correct continuum limit, albeit with a trade--off between discretization and computational cost. Here, we study this interplay for extended reservoir simulations, where relaxation maintains a bias or temperature drop across the system.  Our analysis begins in the non--interacting limit, where we parameterize different discretizations to compare them on an even footing.  For many--body systems, we develop a method to estimate the relaxation that best approximates the continuum by controlling virtual transitions in Kramers' turnover for the current.  While some discretizations are more efficient for calculating currents, there is little benefit with regard to the overall state of the system.  Any gains become marginal for many--body, tensor network simulations, where the relative performance of discretizations varies when sweeping other numerical controls.  These results indicate that a given reservoir discretization may have little impact on numerical efficiency for certain computational tools.  The choice of a relaxation parameter, however, is crucial, and the method we develop provides a reliable estimate of the optimal relaxation for finite reservoirs. 
\end{abstract}

\maketitle
\section{Introduction}

The design of new electronic materials and nanoelectronic devices requires scalable, high--fidelity approaches to simulate transport.   Modern methods can accurately describe the atomic and band structure of many materials, often using density functional theory~\cite{maassen_quantum_2013,kurth_transport_2017,thoss_perspective_2018}.  
Moreover, dedicated many--body techniques, such as quantum Monte Carlo or tensor networks, can include contributions from explicit correlations~\cite{hartle_transport_2015,krivenko_dynamics_2019,ridley_numerically_2019,rams_breaking_2020,wojtowicz_open-system_2020,brenes_tensor-network_2020,lotem_renormalized_2020,fugger_nonequilibrium_2020}. The computational cost of these tools is nonetheless appreciable for large systems or long simulation timescales. These limitations are  particularly onerous for tensor networks, where an explicit treatment of the reservoirs will introduce many degrees of freedom~\cite{dorda_auxiliary_2014,dorda_auxiliary_2015,schwarz_lindblad-driven_2016,fugger_nonequilibrium_2018,rams_breaking_2020,wojtowicz_open-system_2020,brenes_tensor-network_2020,lotem_renormalized_2020,fugger_nonequilibrium_2020}.

\begin{figure}[t!]
\begin{center}
\includegraphics[scale=1.00]{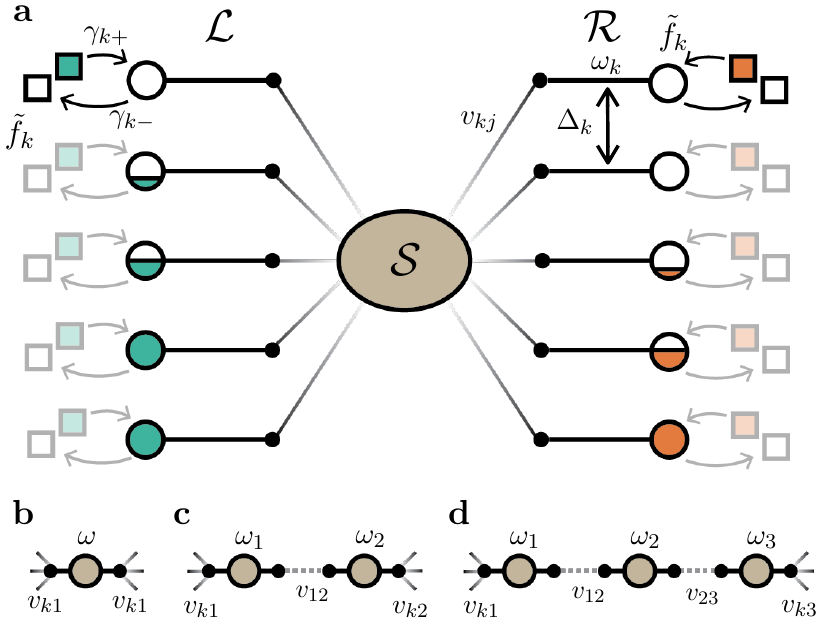}
\end{center}
\caption{{\bf Quantum transport with extended reservoirs.} (a) An arbitrary impurity ($\qs$) is flanked by explicit left ($\ql$) and right ($\qr$) reservoirs.  Each reservoir contains $N_r$ modes of frequencies $\omega_k$ and coupling to a designated system site $i$ with constant strength $v_{ki}$. Implicit reservoirs relax $\ql$ and $\qr$ to biased Fermi distributions $\tilde{f}_k$ at a rate $\gamma_k$ for the $k^\mathrm{th}$ mode. We consider models for $\qs$ that include (b) one, (c) two, and (d) three site systems with onsite frequencies $\omega_{i}$, coupled to each other with strength $v_{ij}$ and to the reservoirs at terminal sites (i.e., a single system mode couples to each reservoir).
}\label{fig:schematic}
\end{figure}

A typical transport simulation is shown in Fig.~\ref{fig:schematic}, where a system (device) of interest is coupled to explicit reservoirs.  Transport is maintained by an external bias. In a closed system, this could be introduced by a density imbalance or a time--dependent, inhomogeneous on--site potential in the reservoirs.  Open systems can go a step further by including implicit reservoirs, which drive transport by relaxing  explicit reservoir modes to biased Fermi distributions~\cite{gruss_landauers_2016,elenewski_communication_2017,gruss_communication_2017,gruss_graphene_2018,zwolak_analytic_2020,zwolak_comment_2020,wojtowicz_dual_2021}. The \emph{extended reservoir approach} exemplifies such an arrangement, and it has become popular in many guises~\cite{gruss_landauers_2016,elenewski_communication_2017,gruss_communication_2017,gruss_graphene_2018,zwolak_analytic_2020,zwolak_comment_2020,kohn_quantum_1957,frensley_simulation_1985,frensley_boundary_1990,mizuta_transient_1991,fischetti_theory_1998,fischetti_master-equation_1999,knezevic_time-dependent_2013,dzhioev_super-fermion_2011,hod_driven_2016,zelovich_state_2014,zelovich_moleculelead_2015,zelovich_driven_2016,zelovich_parameter-free_2017,morzan_electron_2017,ramirez_driven_2019,chiang_quantum_2020,oz_numerical_2020}, including those that accommodate many--body transport~\cite{wojtowicz_open-system_2020,brenes_tensor-network_2020,lotem_renormalized_2020,fugger_nonequilibrium_2020}.

\par  These computational methods require reservoirs that are discretized.  While a given discretization should  converge to the spectral function of a continuum reservoir, its construction is otherwise arbitrary. This flexibility has spawned a variety of methods, including discretizations that place modes evenly across the bandwidth (linear discretization), assign modes from canonical transforms of finite tight--binding lattices, distribute them evenly inside the bias window and logarithmically outside (linear--logarithmic)~\cite{jovchev_influence_2013, schwarz_lindblad-driven_2016, schwarz_nonequilibrium_2018,lotem_renormalized_2020}, or use an influence--based approach to give a linear spacing across the bias window and an inverse spacing outside (linear--inverse)~\cite{zwolak_finite_2008}. Related techniques aim to minimize the number of reservoir modes by introducing {\em intermode} transitions during relaxation.  While these additional fitting parameters can be leveraged to achieve a given level of approximation~\cite{arrigoni_nonequilibrium_2013,dorda_auxiliary_2014,dorda_auxiliary_2015,fugger_nonequilibrium_2018}, they also add long--range couplings which makes tensor network simulations costly. It is unclear which distribution performs best, as a quantitative comparison does not exist.

\par Here, we examine how reservoir parameters---including discretization, system--reservoir coupling, and implicit relaxation---impact the convergence of steady--state transport. We study non--interacting systems and their many--body counterparts, but only consider  extended reservoirs with \emph{intramode} Markovian relaxation~\cite{gruss_landauers_2016,elenewski_communication_2017,gruss_communication_2017,gruss_graphene_2018,wojtowicz_open-system_2020,zwolak_analytic_2020,zwolak_comment_2020,wojtowicz_dual_2021}. For non--interacting systems, we optimize the relaxation (e.g., discretization and coupling to implicit modes) to get the highest accuracy in steady--state currents. This procedure has limited generality since it requires knowledge of the exact, continuum reservoir solution.  For the many--body case, we demonstrate how Kramers' turnover can be used to estimate an optimal relaxation rate.

\par We find that certain discretizations can increase efficiency for non--interacting calculations, where efficiency is measured by the number of reservoir modes required to reproduce the current at a fixed accuracy. This advantage is weak for other system observables (i.e., the impurity's density or correlation matrix), particularly when working at small to moderate reservoir sizes.  While tensor network calculations exhibit moderate, discretization--dependent deviations in the impurity correlation matrix, we find that the overall efficiency is tied to other control parameters---most importantly, the Schmidt cutoff.  This behavior reflects the natural structure of our tensor network, which uses an energy/momentum basis for the isolated reservoirs.  While certain discretizations can mitigate modes that are weakly correlated, these contribute little to the numerical cost.   Thus, the choice of discretization has  little practical impact on efficiency. 

\section{Background and setup}

\par We follow a conventional arrangement~\cite{meir_landauer_1992,jauho_time-dependent_1994} that consists of non--interacting left ($\ql$) and right ($\qr$) reservoirs, and a bias that drives transport through a impurity system ($\qs$), see Fig.~\ref{fig:schematic}.  The associated Hamiltonian has the form $ H = H_\qs + H_\ql + H_\qr + H_\qi$, where $H_\qs$ is the (potentially many--body) Hamiltonian for $\qs$, $H_{\ql(\qr)} = \sum_{k\in \ql(\qr)} \hbar \omega_k \cid{k} \ci{k}$ are the reservoir Hamiltonians, and $H_\qi = \sum_{k\in \ql\qr} \sum_{i\in \qs}  \hbar\, (  v_{ki} \cid{k} \ci{i} +  v_{ik} \cid{i} \ci{k})$ is the interaction Hamiltonian that couples $\qs$ to $\ql\qr$.  The $\cid{m}$ ($\ci{m}$) are fermionic creation (annihilation) operators for a state $m \in \ql\qs\qr$. All indices may implicitly include multiple relevant labels (such as mode number, reservoir, spin). The frequency for the $k^\mathrm{th}$ reservoir mode is denoted by $\omega_k$, while  $v_{ki} = v_{ik}^*$ is used for the coupling between $i \in \qs$ and $k \in \ql\qr$. For two--site impurity $\qs$, the Hamiltonian is 
\begin{eqnarray}
H_\qs &=& \hbar v_\qs (c_1^\dagger c_2 + c_2^\dagger c_1) + \hbar U n_1 n_2,
\end{eqnarray}
where $v_\qs$ is the internal coupling in $\qs$, $n_i = \cid{i} \ci{i}$ is the particle number operator for site $i$, and $U$ is the many--body density--density interaction strength~\cite{wojtowicz_open-system_2020} (the description of other models can be found in the Supplemental Information (SI)).  This model corresponds to a (time--independent) photoconductive molecular device where spin can be neglected~\cite{zhou_photoconductance_2018}. We calculate the properties of non--interacting systems, including the impurity's correlation matrix, using non--equilibrium Green's functions~\cite{gruss_landauers_2016,elenewski_communication_2017,gruss_communication_2017,zwolak_analytic_2020,zwolak_comment_2020}, and employ tensor networks for the many--body case~\cite{wojtowicz_open-system_2020,wojtowicz_dual_2021}.

\par We quantify accuracy of the steady--state current $I$ using a relative error  $| I-I^\circ |/I^\circ$, where the reference current $I^\circ$ is the Landauer limit for  continuum reservoirs (we work with the current itself for many--body cases, as $I^\circ$ is not known exactly).  Furthermore, we quantify combined error in  occupancies and correlations using the correlation matrix of $\qs$, i.e. $\qc_\qs = \qc_{ij} = \langle c_i^\dagger c_j\rangle$, with $i,j \in \qs$.  The quantity $\qc_\qs$ completely characterizes non--interacting systems, and includes the information on densities (occupancies) $n_i = \qc_{ii}$.  A natural metric for convergence of the system state is the normalized trace distance, $\vert\vert C_\qs - C_\qs^\circ \vert\vert_* = \vert\vert C_\qs - C_\qs^\circ \vert\vert / \, 2\,[\tr\, C_\qs + \tr\, C_\qs^\circ]$, defined in terms of the trace norm $\vert \vert M \vert \vert =  \tr \sqrt{M^\dagger M}$ and the correlation matrix $C_\qs^\circ$ for continuum reservoirs.

\par Discretizations are compared by maintaining a common set of modes within the bias window $\qb$, while distributing modes outside the bias window $\qw \setminus \qb$ according to a designated arrangement (here $\qw$ is the reservoir bandwidth). We formalize this by associating an abstract influence function $\zi(\omega)$ with each discretization, and define integrated weights for modes inside $\zI_\qb = \int_\qb \zi(\omega)\, d\omega$ and outside $\zI_{\qw\setminus \qb} = \int_{\qw\setminus\qb} \zi(\omega) \, d\omega$ the bias window.  Similarly, we introduce an influence scale $\zid$ (a target weight per each mode) that gives $N_\qb = \lceil \zI_\qb / \zid \rceil$ modes in the bias window and  $N_{\qw\setminus\qb} = \lceil \zI_{\qw\setminus\qb}  / \zid \rceil$ outside the bias window. The region $\qb$ is then divided into $N_\qb$ bins $\Delta_k$ with boundaries satisfying $\int_{\Delta_k} \zi(\omega)\, d\omega = \zI_\qb/N_\qb$ and $\cup_{k\in\qb} \,\, \Delta_k = \qb$ (similarly for the complement of $\qb$). We choose values of $\zid$ so that there is always an even number of modes in both $\qb$ and $\qw\setminus\qb$.  This accommodation ensures that there is never a mode at the Fermi level.  Reservoir modes are ultimately placed at the midpoint $\omega_k$ of each bin.

\par
\begin{figure}[t]
\begin{center}
\includegraphics[scale=1.0]{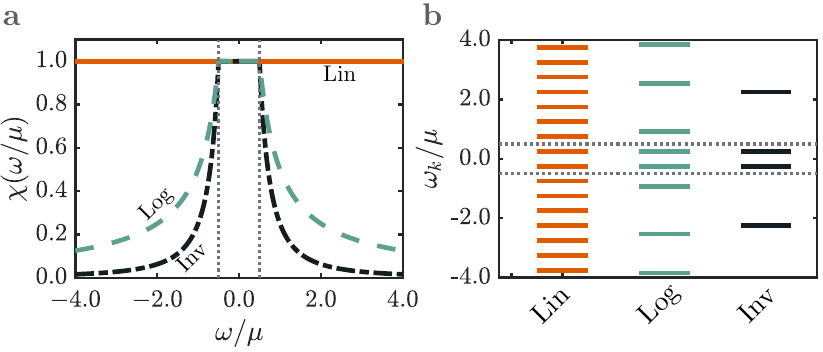}
\end{center}
\caption{{\bf Influence functions and discretizations.} (a) Influence functions that induce linear ($\zi_\text{lin}(\omega)$; orange, solid), linear--logarithmic ($\zi_\text{log}(\omega)$; green, dashed), and linear--inverse ($\zi_\text{inv}(\omega)$; black, dash-dot) discretizations. (b) The resulting minimal mode distributions $\omega_k$, calculated at the same influence scale $\zid$. Thin dotted lines in both plots demarcate the bias window edge. Data are at a bias $\mu = \omega_0 / 2$ and reservoir bandwidth $\qw = 4\,\omega_0$, where $\omega_0$ is the real--space hopping in the reservoir. Modes near the band edges of the linear--logarithmic discretization are a consequence of the chosen influence scale, bias, and bandwidth---they are not necessarily present for denser distributions.} \label{fig:discfunc}
\end{figure}

 We compare three reservoir discretizations: (i) a linear case, with modes spaced evenly throughout the bandwidth; (ii) a linear--logarithmic discretization (motivated by energy scale separation under the numerical renormalization group~\cite{bulla_numerical_2008}); and (iii) a linear--inverse arrangement following the influence approach of Ref.~\cite{zwolak_finite_2008}. The influence functions for these discretizations are 
\begin{eqnarray}
\zi_\text{lin} (\omega) &=& 1 \\
\zi_\text{log}(\omega) &=& \theta\left(\frac{\mu}{2} - \vert \omega \vert\right) + \frac{\mu}{2\vert\omega\vert} \theta\left(\vert \omega \vert - \frac{\mu}{2}\right) \\
\zi_\text{inv}(\omega) &=& \theta\left(\frac{\mu}{2} - \vert \omega \vert\right) + \left(\frac{\mu}{2\omega}\right)^2 \theta\left(\vert \omega \vert - \frac{\mu}{2}\right) ,
\end{eqnarray}
which are nonzero within the reservoir bandwidth and zero outside, as depicted in Fig.~\ref{fig:discfunc}a.  Here,  $\theta(x)$ is the Heaviside step function.  All three measures give evenly spaced modes within $\qb$ yet differ in $\qw \setminus \qb$, acknowledging that bias window modes contribute significantly to the current. Our terminology reflects a measure of influence that is given by the integral of $\zi$.

\par Using these, we compare cases: (i) where the reservoir relaxation is a fixed multiple of the mean level spacing in the bias window $\eta_\text{mean}(\omega_k) = \langle \Delta_k \rangle_{\qb}$ (this is equal to $\mu/N_\qb$ for all cases herein); and (ii) when the relaxation is defined by the mode--dependent level spacing $\eta_\text{level}(\omega_k) = \Delta_k$. We also consider system--reservoir couplings that are defined by  the midpoint between two discrete reservoir modes or by the integrated coupling over an interval of width $\Delta_k$ about a mode $\omega_k$~\footnote{Explicitly, the midpoint coupling for the reservoir mode at $\omega_k$ is derived to match the spectral density in the thermodynamic limit (i.e., reservoirs which are a continuum of states) at the midpoint of an interval $\omega_k \pm \Delta_k / 2$, yielding $v_k = [4 \, v_0^2 \Delta_k \sqrt{1- (2\omega_k / W)^2} / W \pi]^{1/2}$.  Conversely, the integrated coupling maintains the total spectral weight from the continuum reservoirs within the interval $\omega_k \pm \Delta_k / 2$, which gives $v_k = v_0 \pi^{1/2} [K(\omega_k + \Delta_k/2) - K(\omega_k - \Delta_k/2)]^{1/2}$, as defined in terms of the quantity $K(\omega) = 2 \omega (1 - 4\omega^2 / W^2)^{1/2}/W + \csc^{-1} (W / 2\omega)$.  Here, $v_0$ is the system--reservoir coupling in the thermodynamic limit.}.

\section{Kramers' turnover}

\par
\begin{figure}[t]
\begin{center}
\includegraphics[scale=0.95]{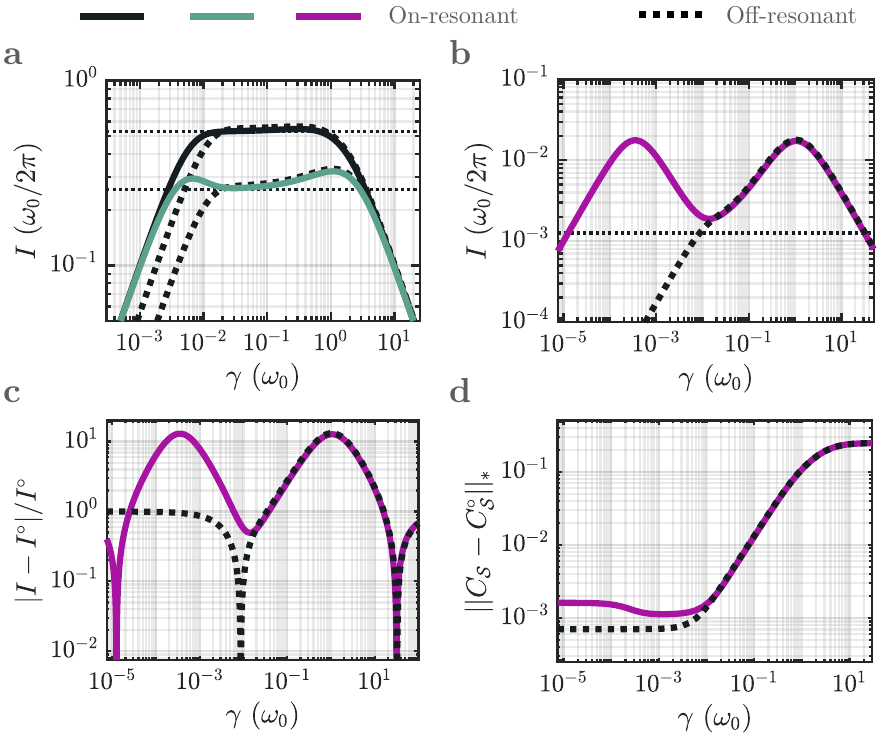}
\end{center}
\caption{ {\bf Kramers' turnover and accuracy}. Steady--state transport for a two--site model with linearly--discretized reservoirs and mode--independent reservoir relaxation $\gamma_k = \gamma$.  Data are presented for on--resonant reservoir modes (solid lines) and those made off--resonant via a frequency shift $\langle \Delta_k \rangle_{\qb} / 2$ between isoenergetic modes (dotted lines).  (a)~Current turnover $I(\gamma)$ at strong system--reservoir coupling $v_0 = \omega_0 / 2$ and two system--site couplings, $v_\qs = (1 + \sqrt{2}) \omega_0 / 4$ (black) and  $v_\qs = (2 + \sqrt{3})\omega_0 / 4$ (green), showing different plateau topographies at different intrasite coupling scales.   (b)~Current turnover at small system--reservoir coupling, $v_0 = \omega_0/10$, revealing anomalies on either side of an interstitial Landauer regime $(v_\qs = (1 + \sqrt{2}) \omega_0 /4)$. (c)~Relative current error with respect to the continuum limit $I^\circ$ for the model in~(b). (d) Convergence of the system state via the normalized trace distance between finite $C_\qs$ and continuum $C_\qs^\circ$ correlation matrices, illustrating that no conditions are uniformly optimal for all observables (the current impacts this convergence in limited manner; see the SI). All calculations use $N_r = 128$ explicit reservoir sites, a bias of $\mu = \omega_0 / 2$ at $T = \omega_0 / 40$, and integrated couplings (see Ref.~\cite{Note3}), and modes spaced evenly between $\pm \qw/2$.  The continuum (Landauer) limit is denoted by dotted horizontal line.
}\label{fig:turnover}
\end{figure}

 The composite $\ql\qs\qr$ system exhibits distinct transport regimes in the presence of relaxation~\cite{gruss_landauers_2016} which mimic Kramers' turnover for chemical reaction rates, see Fig.~\ref{fig:turnover}a~\cite{kramers_brownian_1940} (a similar result holds for thermal transport~\cite{velizhanin_driving_2011,chien_tunable_2013,velizhanin_crossover_2015,chien_thermal_2017,chien_topological_2018}). When relaxation is weak, transport is determined  by the rate at which  particles and holes are replenished in the extended reservoirs. In this regime the current will rise proportionally with $\gamma_k$, analogous to chemical systems where environmental friction controls the equilibration of  reacting species. When the relaxation is strong, phase coherence is suppressed and the current decays as $\gamma_k^{-1}$.  Here, transport emulates reactions where strong friction redirects partially formed products back to the reactants (i.e., recrossings). The intermediate region contains a plateau--like region where the continuum limit current is reproduced, analogous to reactions that are controlled by the  transition state rate.  As we will emphasize later, the system state does not necessarily reflect the exact model on the whole plateau.  The width of the plateau---and convergence to this limit---is dominated by the number and distribution of explicit reservoir modes. The natural transport rate only predominates in the intermediate region~\cite{gruss_landauers_2016}.

\par The formation of the plateau as $N_r \to \infty$ and $\gamma_k \to 0$ (in that order) is sufficient to determine the continuum current, though not all points on the plateau will correspond to a fully converged system state (e.g., local electronic densities).  Moreover, this regime is not guaranteed to be unambiguous. There may be additional features due to the underlying Hamiltonian~\cite{gruss_communication_2017,wojtowicz_open-system_2020} or the presence of specific \emph{anomalies} which exist on either side of the plateau (Fig.~\ref{fig:turnover}ab)~\cite{gruss_landauers_2016,elenewski_communication_2017,wojtowicz_open-system_2020} (see Ref.~\cite{wojtowicz_dual_2021} for details). For large relaxation, a \emph{Markovian anomaly} is associated with an unphysical  broadening of reservoir modes and the lack of a well--defined Fermi level~\cite{gruss_landauers_2016}. This is a direct consequence of Markovian relaxation, which fills a reservoir mode according to its bare frequency $\omega_k$ rather than accounting for its broadening.  Such behavior can lead to zero bias currents in extreme cases~\cite{gruss_landauers_2016}.  These concerns are irrelevant for non--Markovian relaxation, where reservoir modes are properly occupied according their broadened density of states.

\par For weak relaxation, a \emph{virtual anomaly} occurs due to virtual transitions through the system, specifically  between on--resonant $\ql$ and $\qr$ modes.  This leads to excess transport, as previously seen in Refs.~\cite{wojtowicz_open-system_2020,chiang_quantum_2020} and explained in Ref.~\cite{wojtowicz_dual_2021}. The virtual anomaly can be suppressed by shifting the relative energy of $\ql$ and $\qr$ by half the level spacing, $\Delta_k / 2$, disrupting the resonant structure. 
While anomalous regimes can be difficult to distinguish at strong system--reservoir coupling (e.g., $v_0 \approx \omega_0 / 2$), they become prominent when the coupling is weak (e.g., $v_0 \approx \omega_0 / 10$), see Fig.~\ref{fig:turnover}b. 

\par Various factors, including the finite distribution of reservoir modes and the specific Hamiltonian, can influence the turnover architecture (e.g., weak and strong coupling can have a different optimal relaxation~\cite{wojtowicz_dual_2021}). Thus, we need a method that compares discretizations while not placing any given discretization at a disadvantage \emph{a priori}.  We obtain this for non--interacting systems by choosing a  relaxation that most accurately reflects the steady--state current of continuum reservoirs. For many--body cases, we estimate the optimal relaxation.

\section{Optimal relaxation}

\par We can obtain the exact, continuum--limit current of non--interacting systems using established methods.  For finite reservoirs, there is also an optimal relaxation that best estimates this current in the intermediate, physical turnover regime (see Fig.~\ref{fig:turnover}; we exclude incidental crossovers at weak and strong relaxation).  To proceed, we must quantify this optimum for reservoirs with an inhomogeneous mode spacing.  We begin by introducing a relaxation $\gamma_k = \alpha \, \eta(\omega_k)$, where $\alpha$ is a real scaling constant and $\eta( \omega_k)$ is a function of the level spacing within the extended reservoirs.  Using this convention, we can examine cases where $\eta(\omega_k)$ is either: (i) an arbitrary constant; (ii) set equal to the bias window level spacing, which is linearly spaced for the cases we consider; or (iii) set to the $k$--dependent level spacing. We then seek an $\alpha^\star$ in the plateau region that minimizes the relative current error $\alpha^\star = \arg \min (\vert I[\gamma_k(\alpha)] - I^\circ\vert / I^\circ)$ with respect to the continuum limit $I^\circ$. This $\alpha^\star$ completely defines the optimal relaxation for both equally and unequally spaced cases (with a single $\gamma^\star=\gamma_k$ for equally spaced modes).  In principle, we could also derive an optimal relaxation using the normalized trace distance between correlation matrices (see Fig.~\ref{fig:turnover}d) though we do not take this approach. Convergence of this quantity would ensure convergence of all other system observables~\cite{nielsen_quantum_2010}, including the current if there is a boundary that divides the impurity into left and right parts. This relaxation is not required to coincide with $\gamma^\star$ as defined above~\footnote{The current is often only a small contribution to the trace distance. When this is the case, the relaxation that optimizes the trace distance comes at a smaller relaxation strength for the cases we examined.}.

\par It is often impossible to find an optimal $I[\gamma_k(\alpha)]$ for interacting systems since the reference current $I^\circ$ is unknown.  This point is critical in practical calculations.  Optimization can also fail when the plateau is featureless (e.g., at strong--coupling in Fig.~\ref{fig:turnover}a), when many plateau features are present~\cite{gruss_communication_2017}, or if convergence occurs from below the Landauer limit (see the SI).  We can, however, estimate an optimal regime by applying a relative shift of $\langle \Delta_k \rangle_{\qb}/2$ between isoenergetic states in $\ql$ and $\qr$ reservoirs.  That is, we shift the modes in $\ql$ and $\qr$ by plus/minus a quarter of the level spacing. As noted earlier, this eliminates the virtual anomaly associated with resonant transitions~\cite{wojtowicz_dual_2021}. The shifted profile should intersect the unshifted profile at a point $\gamma_s$ near the Landauer regime $\gamma^\star$~\footnote{We can uniquely define $\gamma_s$ only when the two curves intersect. This is the case in all the setups that we study here, however they do not always share a common large--$\gamma$ regime. It is unknown whether the intersection always happens.}. A second estimate is given by extrapolating the linear, small--$\gamma$ regime of the shifted case and finding the point $\gamma_\ell$ where this intersects the unshifted profile. This $\gamma_\ell$ will lie prior to $\gamma^\star$.

\begin{figure}[t]
\begin{center}
\includegraphics[scale=0.95]{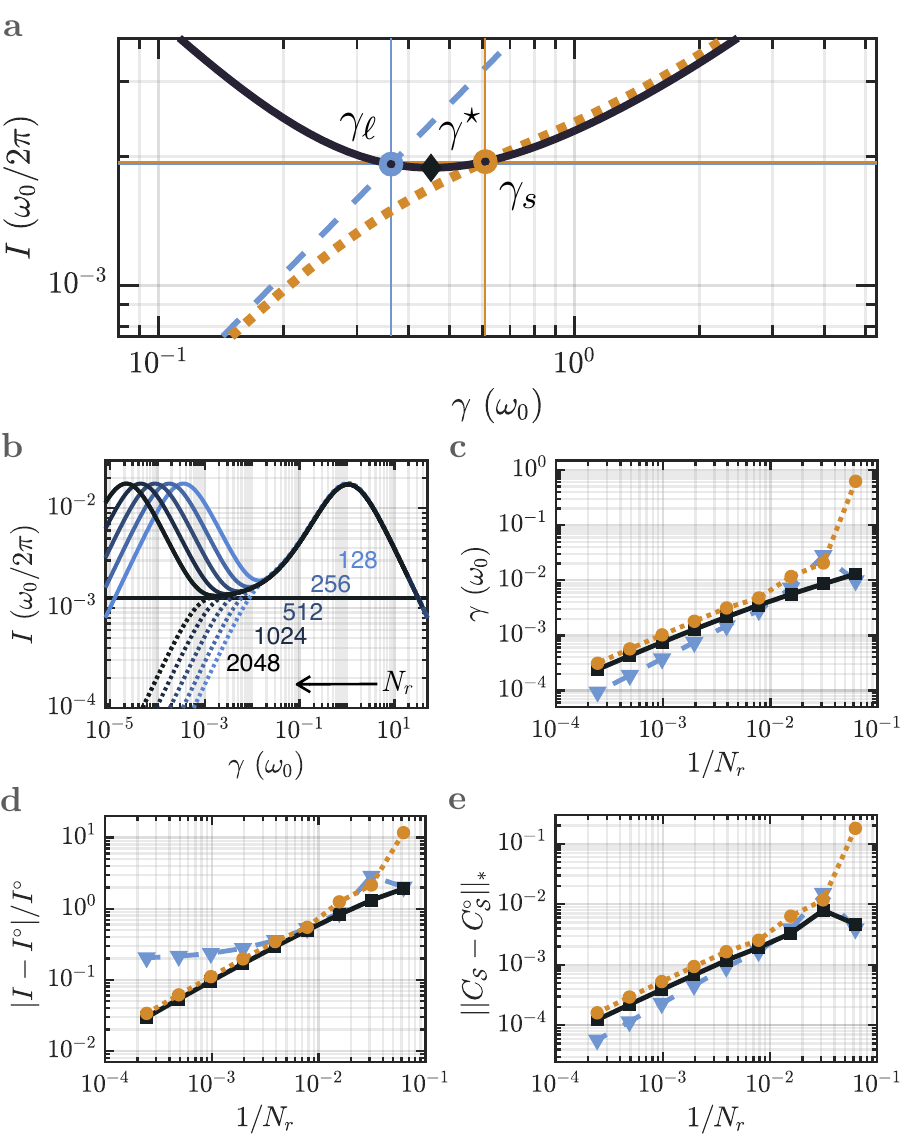}
\end{center}
\caption{{\bf Optimal relaxation and estimators.} (a) Determination of the best estimator for the Landauer regime $I[\gamma_k(\alpha^\star)]$ (black diamond).  Turnover profiles are shown with (tan, dotted line) and without (black, solid line) a frequency shift of $\langle \Delta_k \rangle_{\qb}/2 $ between isoenergetic modes in $\ql$ and $\qr$, defined by the mean level spacing in the bias window.  Estimators are based on linear extrapolation of the small--$\gamma$ regime of the shifted model (off--resonant; blue, dashed) into the unshifted (on--resonant) profile $\gamma_\ell$, or the intersection between shifted and unshifted profiles at $\gamma_s$ (tan circle). (b) Elongation of the region between anomalies as the number of reservoir modes $N_r$ is increased.  (c) Scaling of relaxations associated with $\gamma_\ell$ and $\gamma_s$ estimators. (d)  Convergence of the current error $\vert I - I^\circ\vert / I^\circ$  and (c) the trace norm $\vert\vert C_S - C_S^\circ \vert\vert_*$ with respect to $N_r$. All scaling profiles correspond to  $\gamma_k(\alpha^\star)$  (black, square), the  linear extrapolation estimator (blue, triangle), and the intersection of shifted/unshifted turnover profiles (tan, circle). All panels reflect a linear reservoir discretization for the weak coupling model of Fig.~\ref{fig:turnover}b.} \label{fig:estimators}
\end{figure}

Figure~\ref{fig:estimators}a shows these two estimators. Since the region between anomalies expands into almost flat profile with an increasing number of reservoir sites, we expect these estimators to bound $\gamma^\star$ on either side for large $N_r$. This is indeed the case here. Moreover, the intersection estimator $\gamma_s$ tightly reproduces the optimal point $I[\gamma_k(\alpha^\star)]$ beyond moderate $N_r$.  The placement of reservoir modes plays an a notable role at small--to--moderate $N_r$, especially at weak relaxation, where each mode contributes a narrow peak to the $\ql\qs\qr$ density of states. This underscores the strength of $\gamma_s$ as an estimator, as it lies closer to the large--$\gamma$ regime and thus is less prone to discrepancies from mode placement. 
In contrast, the extrapolation estimator $\gamma_\ell$ is consistently displaced from the physical regime (see Fig.~\ref{fig:estimators}c and the SI). This is a consequence of the plateau topography. That is, the estimator $\gamma_\ell$ scales with $1/N_r$ and rides the edge of the virtual anomaly as $N_r \to \infty$. Hence, its error saturates at a minimum value and it ceases to be a good estimate at large $N_r$. Such behavior is a consequence of the duality between virtual and Markovian anomalies, which can make the optimal relaxation scale as $1/\sqrt{N_r}$ in some regimes~\cite{wojtowicz_dual_2021}.  This saturation does not occur between  $\gamma_\ell$ and the system state, as  $C_\qs$ progressively approaches the continuum limit when increasing $N_r$ at small--to--moderate relaxation (see Fig.~\ref{fig:estimators}d and the SI).

\par The intersection estimator $\gamma_s$ is also robust when examining the overall state of the system (Fig.~\ref{fig:estimators}d).  However, the extrapolation estimator actually outperforms both the optimal and intersection estimators for this case. This is incidental and due to the fact that smaller relaxations often result in a more accurate system correlation matrix, as noted above.  Thus, to find the Landauer limit, we only need to calculate turnover profiles with on--resonant and off--resonant reservoir modes and find their intersection $\gamma_s$---an approach that is borne out for other models and in the strong coupling limit (see the SI). While Hamiltonian parameters can change the plateau architecture, the intersection between turnover profiles will invariably remain a useful estimator of the physical (Landauer) regime.

\section{Results}

Having established a framework to compare different discretizations, we now  examine both non--interacting and many--body transport. As a first step, we compare different system--reservoir coupling methods and different choices of $\eta( \omega_k)$ for the non--interacting case.

\subsection{Non-interacting systems}

\begin{figure}[t!]
\begin{center}
\includegraphics[scale=0.95]{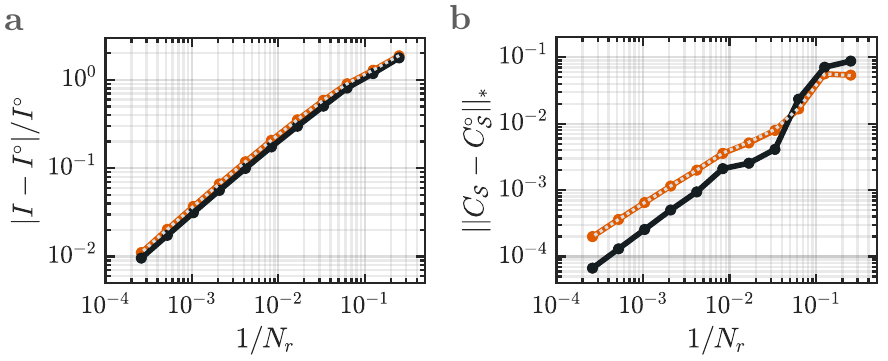}
\end{center}
\caption{{\bf Coupling and relaxation methods.} Convergence of the steady--state current $I[\gamma_k(\alpha^\star)]$ for mode--dependent assignments of reservoir couplings and relaxations. Error in $I$ is shown for (a) integrated couplings with relaxation defined by the mean bias window spacing $\gamma_k = \alpha \langle \Delta_k \rangle_{\qb}$ (black, solid) or by the level spacing $\gamma_k = \alpha \Delta_k$ (orange, dashed). (b) Convergence of the system state, as reflected by the normalized trace distance $\vert\vert C_S - C_S^\circ \vert\vert_*$ between correlation matrices, for the same methods as (a).  They grey dotted line in (a) and (b) reflects $\gamma_k$ based on a level--dependent bias window spacing, but with couplings from the midpoint of the discretization intervals.  The model is otherwise that of Fig.~\ref{fig:turnover}b with a linear--inverse discretization.}\label{fig:methodCompare}
\end{figure}

\par The behavior of a reservoir discretization may be influenced by the system--reservoir coupling and the assignment of relaxation rates $\gamma_k$ to each reservoir mode.  We present this behavior for the linear--inverse discretization in Fig.~\ref{fig:methodCompare}.  The most significant factors are the relaxation rates, which nontrivially moderate convergence to the continuum limit with increasing $N_r$.  The error in $I$ is minimized when the relaxation is a multiple of the mean level spacing in the bias window, $\gamma_k = \alpha \langle \Delta_k \rangle_{\qb}$ (which, in the cases here, is equal to $\mu/N_\qb$). This situation is more variable for convergence of $C_\qs$, where we see better performance at small $N_r$ if the relaxation is a multiple of the level spacing $\gamma_k = \alpha \Delta_k$ (Fig.~\ref{fig:methodCompare}a,b).  Nonetheless, this behavior crosses over to favor the mean--spacing approach at modest $N_r$.   We note that convergence is minimally impacted by the coupling method---the integrated and mean methods do not differ  appreciably at any $N_r$ scale.

\begin{figure}[t!]  
\begin{center}
\includegraphics[scale=0.95]{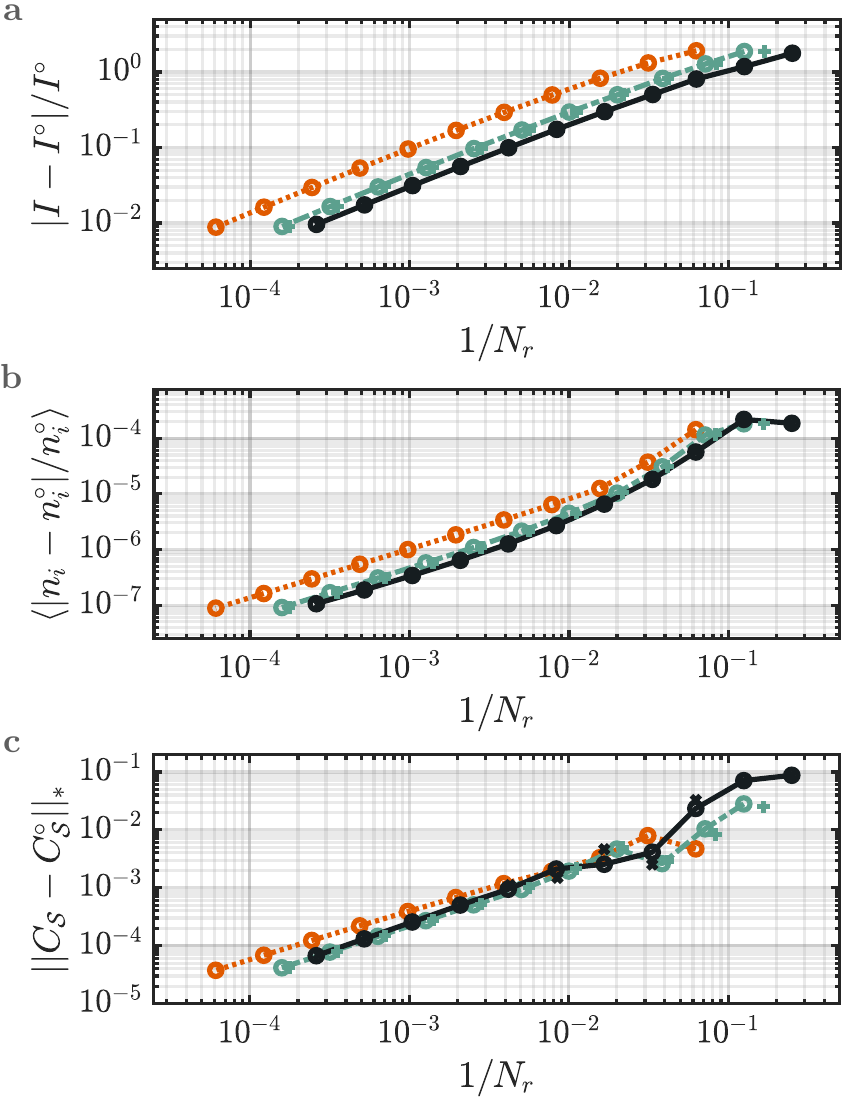}
\end{center}
\caption{{\bf Error and discretization.} Convergence of reservoir discretizations when  increasing the number $N_r$ of explicit reservoir sites.  This behavior is quantified through (a) relative error in the steady--state current $I[\gamma_k(\alpha^\star)]$; (b) the mean relative error of the on--site densities $n_i$ within $\qs$; and (c) the normalized trace distance between correlation matrix $C_\qs$ for $\qs$ and its infinite reservoir counterpart $C_\qs^\circ$. Discretizations correspond to the standard linear (orange, dotted line), the linear--logarithmic (green, dashed line), and the linear--inverse (black, solid line) arrangements.  Results are also provided for additional linear--logarithmic and linear--inverse discretizations which are the transform of a 1--d spatial lattice to the energy basis (green and black crosses).  Profiles from (a) fit to $A/N_r^p$ with $[A,p] = [11 \pm 1, -0.65 \pm 0.02] $, $[8.5 \pm 0.2, -0.72 \pm 0.01]$, and $[4.4 \pm 0.2 , -0.64 \pm 0.18 ]$ for the main  discretizations.   All data are from the non--interacting, two--site Hamiltonian of Fig.~\ref{fig:turnover}b at weak--coupling ($v_0 = \omega_0 / 10$), with integrated system--reservoir couplings, and relaxations $\gamma = \alpha \langle \Delta_k\rangle_{\qb}$ determined by the mean mode spacing within the bias window $\mu = \omega_0 / 2$.}
\label{fig:influenceCompare}
\end{figure}

\begin{figure}[t!]  
\begin{center}
\includegraphics[scale=0.95]{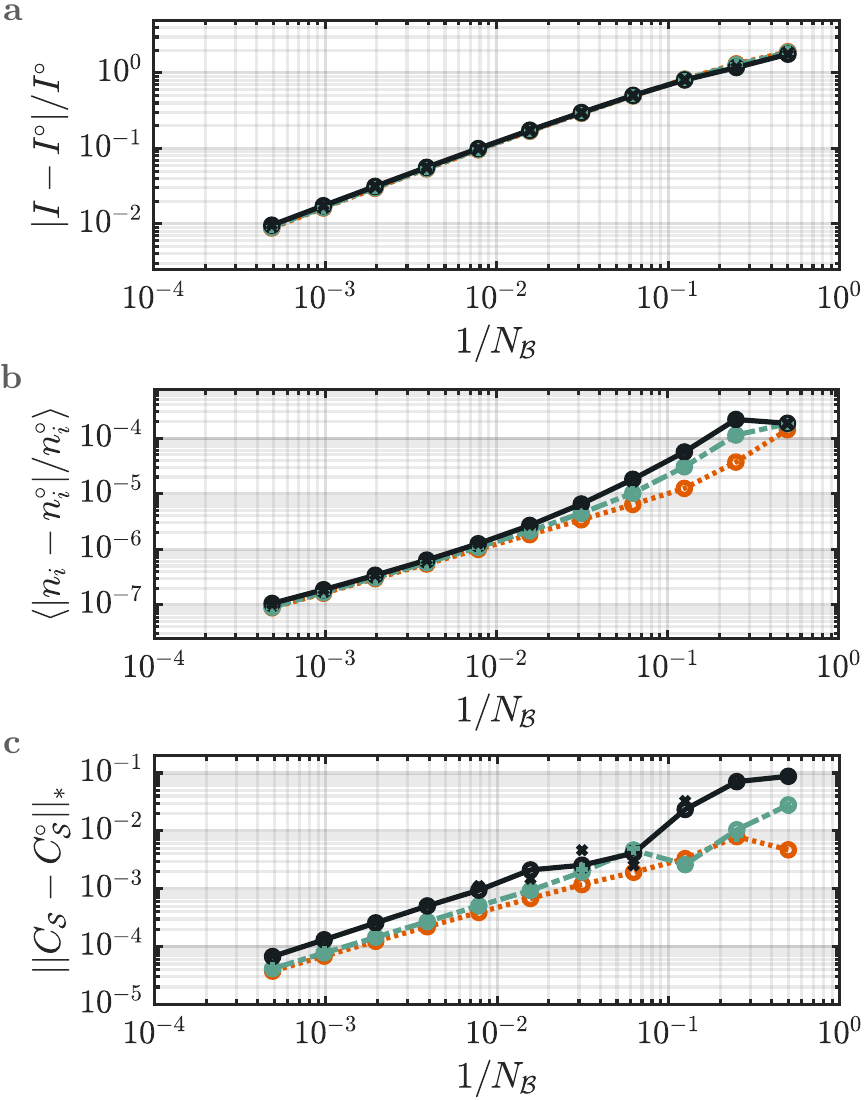}
\end{center}
\caption{{\bf Error and discretization in the bias window.} Convergence of the reservoir discretizations from Fig.~\ref{fig:influenceCompare}, now parameterized in terms of the number of states $N_\qb$ in the bias window.  Scaling is quantified through (a) relative error in the steady--state current $I[\gamma_k(\alpha^\star)]$; (b) the mean relative error of the on--site densities $n_i$ within $\qs$; and (c) the normalized trace distance between correlation matrix $C_\qs$ for $\qs$ and its infinite reservoir counterpart $C_\qs^\circ$.   Colors and symbols follow from Fig.~\ref{fig:influenceCompare}.  Profiles from (a) fit to $A/N_\qb^p$ with $[A,p] = [4.1 \pm 0.1, -0.76 \pm 0.02] $, $[4.0	 \pm 0.2, -0.76 \pm 0.02]$, and $[3.8 \pm 0.3 , -0.74 \pm 0.03 ]$ for the main discretizations, while restricting to $N_\qb > 4$ to mitigate finite size effects.  }
\label{fig:influenceCompareBW}
\end{figure}

\par Figure~\ref{fig:influenceCompare} shows the performance of different discretizations when converging a transport calculation. We find the full linear discretization $\zi_\text{lin}(\omega)$ to behave more poorly than other measures when using either the relative error in current $I$ or in the system--site density $n_i$  as a metric for convergence (Fig.~\ref{fig:influenceCompare}a,b).  Notably, the error in the steady--state current is uniformly higher than other measures at comparable scales of influence for all values of $N_r$.   Using the same criteria, the linear--inverse influence measure $\zi_\text{inv}(\omega)$  outperforms the linear--logarithmic discretization $\zi_\text{log}(\omega)$.  This implies a lower degree of error at fewer reservoir sites, providing better convergence in a regime with  decreased computational cost. The performance gain when moving between these methods is nonetheless smaller than the gain when moving to them from the full linear discretization.  

\par Any seeming advantage is less clear--cut for the overall state of the system, where all three discretizations exhibit comparable performance at large $N_r$. Nonetheless, the linear--inverse discretization performs more poorly when $N_r$ is small---a region where convergence can oscillate due to the placement of states outside the bias window edge. Similar conclusions may be drawn for models containing one or three sites (Fig.~\ref{fig:schematic}b,d; see SI). Such deviations are largely academic, as these methods are roughly equivalent for the maximal number of states used in typical many--body transport simulations (i.e., $N_r$ in the 10's to 100's).

\par A similar analysis can be done in terms of the number of reservoir modes $N_\qb$ within the bias window (Fig.~\ref{fig:influenceCompareBW}).  This region is particularly important when representing the current, and the accuracy of a representation correlates with $N_\qb$.  Working from this perspective, we find uniform scaling across discretizations with respect to the current error.  This observation simply reflects that transport is dominated by bias window modes.  The occupations also scale uniformly at large $N_\qb$, albeit with discrepancies when this parameter is small.  Correlation matrices have more sporadic behavior, though the linear--inverse arrangement reproduces the system state most poorly at a given $N_\qb$.  This is expected since it has the largest percentage of bias window modes  and thus fails to capture correlations elsewhere in the bandwidth.  The performance gap for the linear--inverse is nonetheless offset by the overall reduction in $N_r$ at a given influence scale.

\subsection{Many-body impurities}

Sophisticated numerical methods, such as tensor networks, are required to study complex, interacting models.  We adopt a typical approach for open quantum systems, where the density matrix is vectorized and approximated as a matrix product state (MPS)~\cite{zwolak_mixed-state_2004,verstraete_matrix_2004}.  This construction may be represented diagrammatically as:

\begin{equation}
\includegraphics[width=\columnwidth]{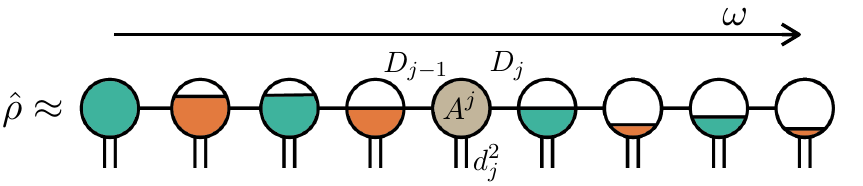}
\label{eq:mps}
\end{equation}
\vspace{0.25\baselineskip}

\noindent where we have ordered the combined $\ql\qr$ modes (green/orange) according to their energies, reflecting the resonant nature of the current-carrying states~\cite{rams_breaking_2020,wojtowicz_open-system_2020} (the color--coding follows Fig.~\ref{fig:schematic}).  The system $\qs$ (grey) is positioned in the middle at $\omega=0$. Following this notation, $d_j$ is the local Hilbert space dimension at site $j$ and $D_j$ is the MPS bond dimension to the right of site $j$. The latter determines the size  $D_{j-1}\times d_j^2\times D_j$ of each tensor $A^j$ constituting the MPS. The computational cost will depend on both $N_r$ and the structure of the correlations, which set the minimal $D_j$ needed to reach a given level of accuracy.  Our choice of reservoir mode ordering has been shown to minimize this bond dimension by mitigating the spread of entanglement~\cite{rams_breaking_2020,wojtowicz_open-system_2020}.  We obtain steady--states by using the time--dependent variational principle~\cite{haegeman_unifying_2016} to evolve an MPS under the Lindblad superoperator, as described in Ref.~\cite{wojtowicz_open-system_2020} (see Ref.~\cite{brenes_tensor-network_2020} for a similar approach with a different state ordering). Since the accuracy of this approach depends on the bond dimension, we can adjust the latter using a cutoff $\epsilon_{\rm min}$.  That is, we only retain the singular values that are above this cutoff for each bipartition of the chain in Eq.~\eqref{eq:mps}.

\begin{figure}[t!]
    \centering
    \includegraphics[width=\columnwidth]{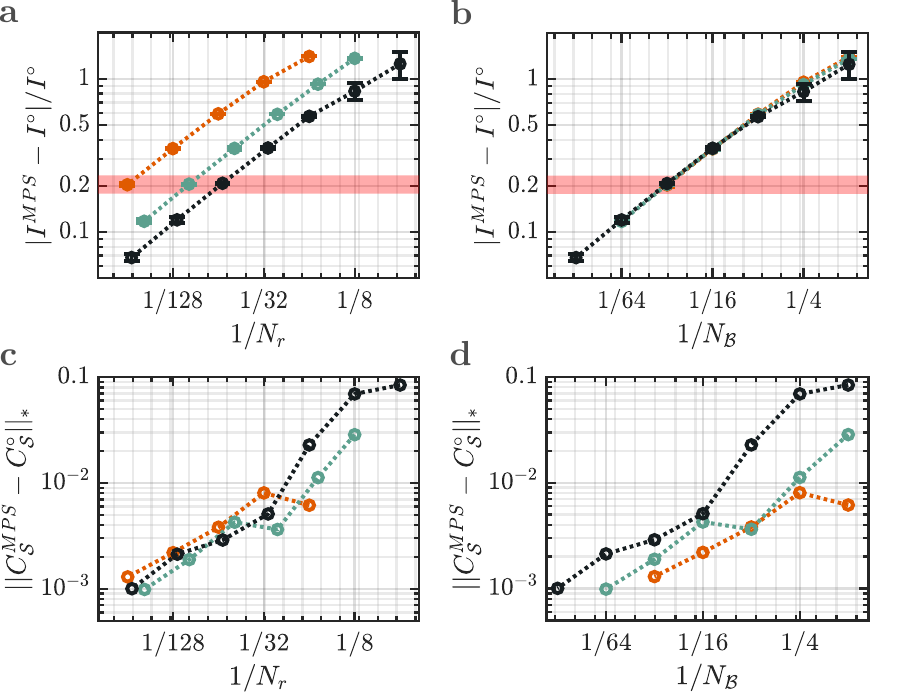}
    \caption{{\bf Error and discretization for non--interacting MPS.} Convergence of reservoir discretizations for the non--interacting two--site system $\qs$ of Fig.~\ref{fig:influenceCompare}, obtained using MPS with a fixed Schmidt cutoff $\epsilon_{min}=10^{-6}$.  Scaling is quantified with respect to the number of modes in each reservoir $N_r$ and the number within each bias window $N_\qb$.  Data correspond to (a,~b)  relative error in the steady--state current $I^{MPS} = I[\gamma^\star]$ versus the Landauer limit $I^\circ$ and (c,~d) the normalized trace distance between correlation matrix $C_\qs^{\rm MPS}$ for $\qs$ and its infinite reservoir counterpart $C_\qs^\circ$. Discretizations follow linear (orange), the linear--logarithmic (green), and the linear--inverse (black) arrangements.   The red band in (a,~b) is a relative error scale (0.20), for which $N_r$ is  256, 100, and 60, respectively.  The current $I^{\rm MPS}$ is an average from $\ql\qs$, $\qs_1\qs_2$, and $\qs\qr$ interfaces.  Uncertainties $\sigma = \pm\sqrt{\sigma_1^2+\sigma_2^2}$ reflect fluctuations $\sigma_1$ of the current over a temporal window $\Delta t = 50\, \omega_0^{-1}$, as well as the mismatch $\sigma_2^2 = \sum_j |I_j-I^{\rm MPS}|^2/3$ of currents at the interfaces $j \in \{\ql\qs_1,\, \qs_1\qs_2,\, \qs_2\qr\}$.  The designated $C_\qs^{MPS}$ is representative of the final simulation time step. Parameters are identical to Fig.~\ref{fig:influenceCompare}, but with a system--reservoir coupling $v_0=\omega_0/8$. }
    \label{fig:mpsA}
\end{figure}

\begin{figure}[t!]
    \centering
    \includegraphics[width=\columnwidth]{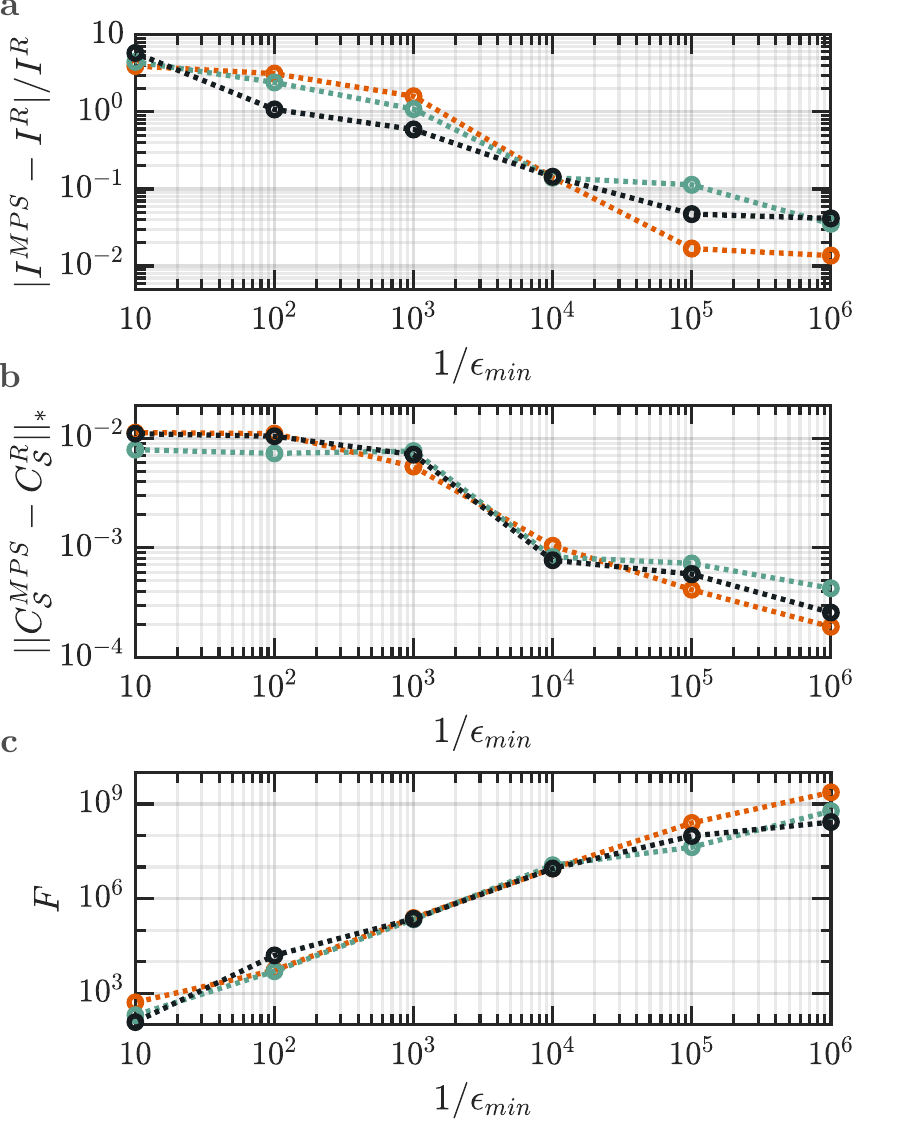}
    \caption{
    {\bf Error and Schmidt cutoff for non--interacting MPS.} 
    Convergence of reservoir discretizations for the non--interacting two--site MPS calculations of Fig.~\ref{fig:mpsA}, now at a fixed error level with respect to the continuum current $I^\circ$ (denoted by the red band in Fig.~\ref{fig:mpsA}a,b).  This behavior is quantified in terms of the Schmidt cutoff $\epsilon_{min}$ for (a) relative error in the steady--state current $I^{MPS}$ versus its exact counterpart $I^R$; (b) the normalized trace distance between correlation matrix $C_\qs^{\rm MPS}$ for $\qs$ and its exact counterpart $C_\qs^R$; and (c) the relative numerical cost of a single MPS update $F=\sum_j D^3_j$, defined in terms of the MPS bond dimensions $D_j$ at all bipartitions. Discretizations correspond to linear (orange), the linear--logarithmic (green), and the linear--inverse (black) arrangements. }
\label{fig:mpsB}
\end{figure}

\begin{figure}[t!]
    \centering
    \includegraphics[width=\columnwidth]{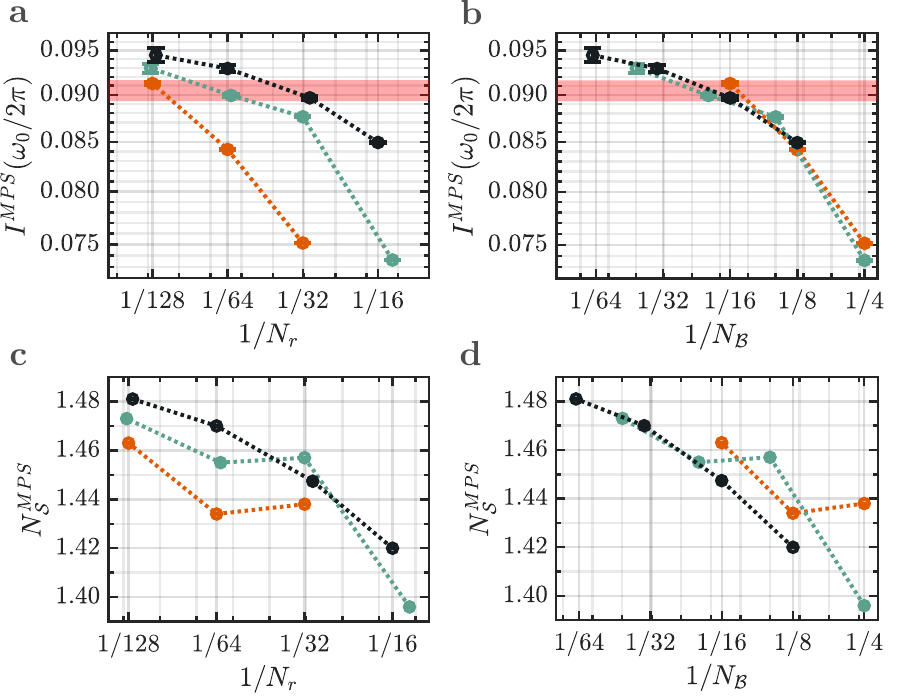}
    \caption{{\bf Error and discretization for interacting MPS.}     Convergence of reservoir discretizations for the two--site MPS calculations of Fig.~\ref{fig:mpsA} with an additional many--body interaction $U=-\omega_0/2$.   Scaling is quantified with respect to the number of modes in each reservoir $N_r$ and the number within each bias window $N_\qb$.  Data correspond to (a,~b)  relative error in the steady--state current $I^{MPS} = I[\gamma^\star]$ versus the Landauer limit $I^\circ$ and (c,~d) the normalized trace distance between correlation matrix $C_\qs^{\rm MPS}$ for $\qs$ and its infinite reservoir counterpart $C_\qs^\circ$. Discretizations follow linear (orange), the linear--logarithmic (green), and the linear--inverse (black) arrangements.   The red band in (a,~b) is a relative error scale (0.20), for which $N_r$ is 128, 62, and 30, respectively.  Model parameters and  uncertainties in the current are identical to Fig.~\ref{fig:mpsA}.  Uncertainties for $N_\qs$ are given by $\sigma = \pm\sigma_1$, reflecting fluctuations of the measurement over a temporal window $\Delta t = 50\, \omega_0^{-1}$.  Calculations reflect MPS with $\epsilon_{min}=10^{-6}$. }
\label{fig:mpsU}
\end{figure}

\begin{figure}[t!]
    \centering
    \includegraphics[width=\columnwidth]{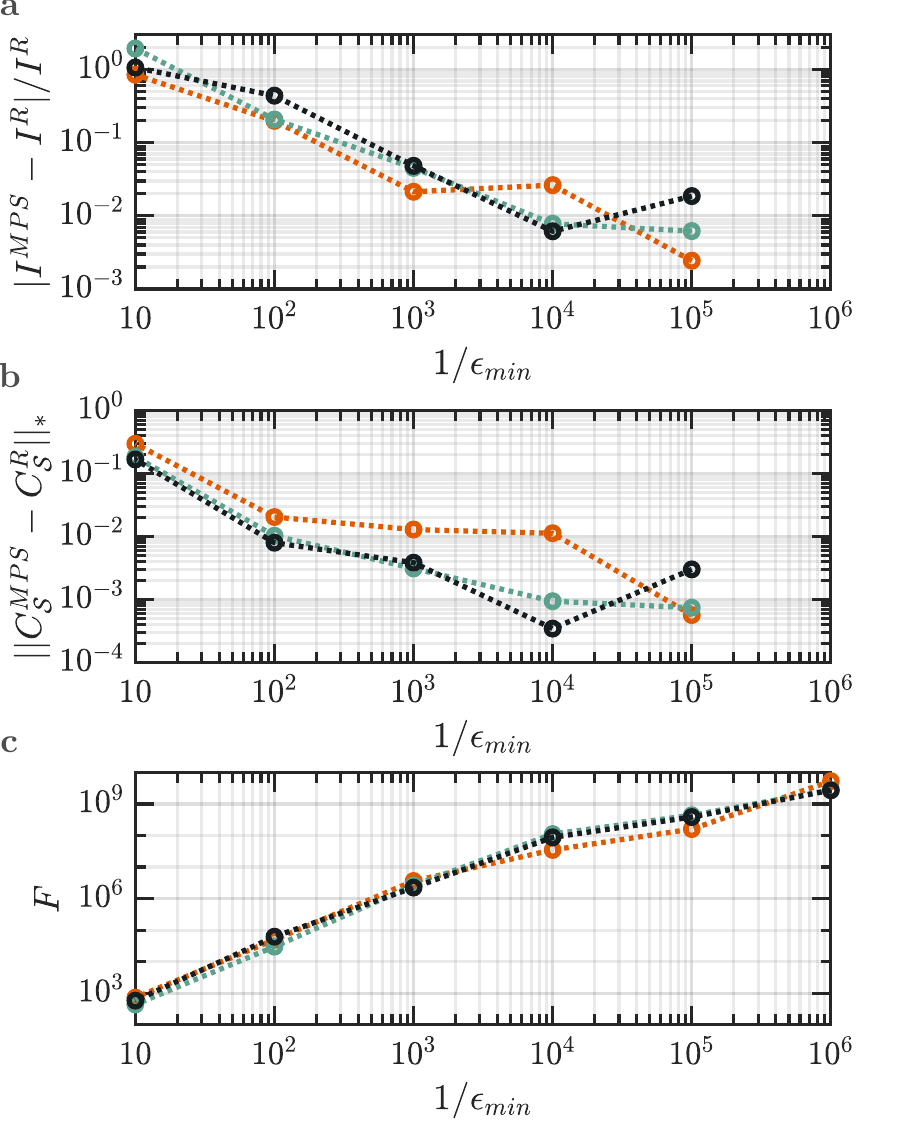}
    \caption{
    {\bf Error and Schmidt cutoff for interacting MPS.} 
       Convergence of reservoir discretizations for the interacting two--site MPS calculations of Fig.~\ref{fig:mpsU}, now at a fixed error level with respect to the continuum current $I^\circ$ (denoted by the red band in Fig.~\ref{fig:mpsU}a,b).  This behavior is quantified in terms of the Schmidt cutoff $\epsilon_{min}$ for (a) relative error in the steady--state current $I^{MPS}$ versus its most converged value $I^R=I^{MPS}[\gamma_s]$ when $\epsilon_{min}=10^{-6}$; (b) the normalized trace distance between correlation matrix $C_\qs^{\rm MPS}$ and its counterpart estimated at $\gamma_s$ and $\epsilon_{min}$ from (a); and (c) the relative numerical cost of a single MPS update $F=\sum_j D^3_j$, defined in terms of the MPS bond dimensions $D_j$ at all bipartitions. Discretizations correspond to linear (orange), the linear--logarithmic (green), and the linear--inverse (black) arrangements.}
    \label{fig:mpsUb}
\end{figure}

\par We quantify convergence of our MPS calculations using the steady--state current, which is consistently larger than other error measures.  Our analysis will focus on  the weakly--coupled, two--site impurity model from Fig.~\ref{fig:influenceCompare} in both non--interacting and interacting limits.  To assess the consistency of our methods, we first confirm that the current and correlation matrix from the non--interacting MPS can reproduce the exact solution for all three discretizations (Fig.~\ref{fig:mpsA}).   This confidence allows us to focus on a particular level of discretization--related error, indicated by the red band in Fig.~\ref{fig:mpsA}a.  By fixing the number $N_r$ of sites in each reservoir to a value within this band, we can determine how the singular value threshold $\epsilon_{\rm min}$ controls convergence of the current and the system state at a given accuracy.  This accommodation also fixes the number of bias window sites $N_\qb$ to be the same for each discretization---an important point that we will address later.  To proceed, we measure error with respect to the exact, finite--size current $I^R$ associated with a given $N_r$ and discretization of a non--interacting system.  We find a numerical solution that slowly approaches the exact current as $\epsilon_{\rm min}$ is decreased, however, this convergence is not uniform (Fig.~\ref{fig:mpsB}a). The choice of discretization has little impact on convergence even though the number of MPS sites is quite different.

\par This behavior can be understood by using the quantity $F = \sum_j D_j^3$ to estimate relative cost of MPS simulations for a given  $\epsilon_{\rm min}$.  This metric encapsulates the scaling of computational time with bond dimension, as other parameters contributing to the cost (e.g., bond dimensions for the Lindbladian MPO, local Hilbert space dimensions) are the same for all discretizations. Our discretizations differ in the total number of reservoir sites $N_r$ that are needed to reproduce a given level of accuracy.  However, an analysis based on $F$ suggests that the degree of correlation is determined by the number of states within the bias window $N_\qb$, which is the same for each discretization at a given accuracy level (Fig.~\ref{fig:mpsB}).   Thus, we cannot specify a   discretization that will yield a clear increase in computational performance for MPS simulations. The only benefit to having a smaller $N_r$ is having fewer modes outside the bias window. This has little computational impact, as our ordering places these modes at corners of the MPS, where they require a small $D_j$ and contribute weakly to $F$.

\par A related analysis can be performed for interacting systems, which we demonstrate by introducing a density--density interaction of strength $U=-\omega_0/2$ between the impurity sites.  Since the exact solution is unknown, we estimate an optimal relaxation $\gamma_s$ by comparing $\gamma$--dependent turnover profiles with on/off--resonant modes (Fig.~\ref{fig:estimators}), as validated earlier in the manuscript.  This procedure is executed for each discretization and set of reservoir modes, yielding the scaling behavior presented in Fig.~\ref{fig:mpsU}a.  We again find a current that converges monotonically with increasing $N_r$ for all discretization schemes, though the convergence of occupations varies more.

\par We can also assess how simulation performance scales with $\epsilon_{\rm min}$ when interactions are present.  Following our analysis for the non--interacting MPS, we define a fixed level of discretization--related error (the red band in Fig.~\ref{fig:mpsU}a, corresponding to $N_\qb = 16$), measured with respect to the limiting, finite--size current $I^R$  for an interacting system.  To avoid finite size effects, we limit this and subsequent analysis to points with $N_\qb > 4$.  We find that convergence of the current and correlation matrix $C_\qs$ is comparable across discretizations, as is the numerical cost quantified through $F$ (Fig.~\ref{fig:mpsUb}c).   Once again, performance is dictated by how accurately we represent the bias window (and thus by $N_\qb$), emulating the non--interacting MPS.  The reservoir discretization still has little impact when converging the current in tensor network simulations  at practical reservoir sizes.  In fact, the Schmidt cutoff $\epsilon_{\rm min}$ and underlying system Hamiltonian  are the primary determinants of convergence.
 s

\par The exact, continuum--limit current is unknown for many interesting systems.    Nonetheless, our extended reservoir simulations should approach this regime as the number of explicit reservoir modes is increased.  This is particularly true for the current, where we have observed monotonic convergence with $N_r$ in both non--interacting and interacting MPS simulations.  We can test this assumption by fitting a scaling law $I = I^\infty + A/N_\qb^\alpha$ to the non--interacting data of  Fig.~\ref{fig:mpsA}, and extract an estimate for the  current $I^\infty$ with continuum reservoirs.  The importance of bias window modes is acknowledged by parameterizing in terms of $N_\qb$.  In this case, we obtain scaling exponents of $\alpha = $ [$0.48 \pm 0.05$, $0.54 \pm0.04$, $0.67 \pm 0.02$] and continuum limit currents of $2\pi I^\infty / \omega_0 =$ [$0.0023 \pm 0.0003$, $0.0026 \pm 0.0002$, $0.0030 \pm 0.0001$]  for the fully linear, linear--logarithmic and linear--inverse discretizations, respectively.  These exhibit reasonable agreement with their exact counterpart  $2\pi I^\circ / \omega_0 = 0.0031$, albeit with some discrepancies.  The high performance of the linear--inverse arrangement is expected since bias window modes predominate for this discretization.

\par Our scaling exponents $\alpha$ can be compared to exact profiles such as Fig.~\ref{fig:influenceCompare}, where we are guaranteed that $I^\infty$ will equate to $I^\circ$ at large $N_r$.  Performing this exact analysis when $v_0 = \omega_0 / 8$ gives scaling exponents of $\alpha = $ [$0.77 \pm 0.01$, $0.77 \pm0.01$, $0.74 \pm 0.01$]. The discrepancy between our MPS fits and the exact result suggests that $\alpha$ is difficult to determine from small $N_r$ data, and that it can vary across different scales of $N_\qb$.  In particular, we see that our fitting procedure gives $\alpha \approx 3/4$.  For increasingly dense mode distributions, we expect that the bias window modes will become dominant and those outside will be marginalized.  This would lead to values of  $\alpha$ that become increasingly homogeneous across discretizations.  If we perform fits by aggregating data from all discretizations, we find an $\alpha = 0.76 \pm 0.01$ for exact simulations.  We likewise obtain $\alpha= 0.72 \pm 0.03$ and $2\pi I^\infty /\omega_0 = 0.0030 \pm 0.0001$ by doing the same for our non--interacting MPS calculations.  This result is closer to expected values.  The same strategy can be applied the interacting system of Fig.~\ref{fig:mpsU}.  Since we have a very limited dataset and no analytical solution for the continuum limit, we forgo analysis in terms of individual discretizations and instead fit the aggregate profile  to find  $\alpha = 1.50 \pm 0.63$ and $2\pi I^\infty/\omega_0 = 0.093 \pm 0.002$.  The large standard error in the exponent may indicate that modes outside the bias window have a greater influence when interactions are present.

\section{Conclusions}

\par Our observations suggest a general approach when using discrete reservoirs in quantum transport simulations.  In a technical sense, we find that the linear--inverse discretization is the most efficient arrangement, particularly when combined with a relaxation method based on the level spacing in the bias window.  Nonetheless, the performance between discretizations is not dramatic, and is effectively negligible for the $N_r$ used in practical simulations.   This is especially true for interacting MPS--based simulations, where correlations ultimately regulate the computational cost.   Despite this behavior, one should remain mindful of cases where the choice of discretization can become more important---notably for small $N_r$ or at a small bias where a large portion of the bandwidth becomes less relevant (at least for the current).  Furthermore, there may remain some interplay between the performance of a given discretization toward a particular observable and the precise distribution of states within $\qs$.  This consideration could be relevant in computationally taxing cases,  including certain many--body limits, where $N_r$ is strongly limited by practical constraints.

\par In addition, we developed a method for estimating the optimal relaxation $\gamma^\star$ that approximates the continuum result $I(\gamma^\star) \approx I^\circ$ at a given scale.  This is especially valuable when the continuum limit $I^\circ$ is unknown.  While the turnover region will vary between model Hamiltonians and coupling regimes, we need only ``switch on'' a level shift between reservoirs and use the intersection $\gamma_s$ between shifted and unshifted turnover profiles (or $\gamma_\ell$ from linear extrapolation) to estimate the best relaxation. This provides a practical tool for performing extended reservoir simulations with matrix product states and tensor networks.

\section{Acknowledgements}

J.~E.~E. acknowledges support under the Cooperative Research Agreement between the University of Maryland and the National Institute for Standards and Technology Physical Measurement Laboratory, Award 70NANB14H209, through the University of Maryland. We acknowledge support by the National Science Center (NCN), Poland under Projects No. 2016/23/B/ST3/00839 (G.~W.) and No. 2020/38/E/ST3/00150 (M.~M.~R.).

\end{document}

\newcommand{\todo}[1] {{\color{red} #1}}

\renewcommand{\theequation}{S\arabic{equation}}
\renewcommand{\thefigure}{S\arabic{figure}}

\global\long\def\avg#1{\langle#1\rangle}
\global\long\def\p{\prime}

\global\long\def\ket#1{|#1\rangle}
\global\long\def\bra#1{\langle#1|}
\global\long\def\proj#1#2{|#1\rangle\langle#2|}
\global\long\def\inner#1#2{\langle#1|#2\rangle}
\global\long\def\tr{\mathrm{tr}}

\global\long\def\rsd#1#2{\frac{\partial^{2}#1}{\partial#2^{2}}}
\global\long\def\der#1#2{\frac{d#1}{d#2}}
\global\long\def\im{\imath}

\newcommand{\<} {\left\langle}
\renewcommand{\>} {\right\rangle}
\newcommand{\dg} {{\dagger}}
\newcommand{\pd} {{\phantom\dagger}}

\newcommand{\ci}[1] {{c_{#1}^{\pd}}}
\newcommand{\cid}[1] {c_{#1}^\dg}

\newcommand{\bi}[1] {{c_{#1}^{\pd}}}
\newcommand{\bid}[1] {c_{#1}^\dg}

\newcommand{\ai}[1] {a_{#1}^\pd}
\newcommand{\aid}[1] {a_{#1}^\dg}
\newcommand{\jinj} {J_{\text{inj}}(\omega)}
\newcommand{\jdepl} {J_{\text{depl}}(\omega)}
\newcommand{\jinje} {J_{\text{inj}}(\epsilon_i)}
\newcommand{\jdeple} {J_{\text{depl}}(\epsilon_i)}

\newcommand{\ginj}[2] {\gamma_{#1^#2}}
\newcommand{\gk}[1] {\ginj{k}{#1}}
\newcommand{\gi}[1] {\ginj{i}{#1}}
\newcommand{\gj}[1] {\ginj{j}{#1}}
\newcommand{\hc} {\text{h.c.}}
\newcommand{\emax} {{\epsilon_{\text{max}}}}
\newcommand{\emin} {{\epsilon_{\text{min}}}}

\newcommand{\bG} {\bm{G}}
\newcommand{\bS} {\bm{\Sigma}}
\newcommand{\bGa} {\bm{\Gamma}}

\renewcommand{\Re} {\operatorname{Re}}
\renewcommand{\Im} {\operatorname{Im}}

\newcommand{\coup}{v_k}

\newcommand{\ql}{\mathcal{L}}
\newcommand{\qs}{\mathcal{S}}
\newcommand{\qr}{\mathcal{R}}
\newcommand{\qe}{\mathcal{E}}
\newcommand{\qi}{\mathcal{I}}
\newcommand{\qu}{\mathcal{U}}
\newcommand{\qc}{\mathcal{C}}
\newcommand{\qb}{\mathcal{B}}

\newcommand{\rs}{\bar{\rho}}
\newcommand{\EO}{E}
\newcommand{\EC}{\mathcal{J}}

\newcommand{\zi}{\chi}
\newcommand{\zI}{X}
\newcommand{\zid}{x}

\newcommand\trick[1]{}

\title{Supplementary Information for ``Performance of Reservoir Discretizations in Quantum Transport Simulations''}
\author{Justin E. Elenewski} 
\affiliation{Biophysical and Biomedical Measurement Group, Microsystems and Nanotechnology Division, Physical Measurement Laboratory, National Institute of Standards and Technology, Gaithersburg, MD, USA}
\affiliation{Institute for Research in Electronics and Applied Physics, University of Maryland, College Park, MD, USA}
\author{Gabriela W\'{o}jtowicz}
\affiliation{Jagiellonian University, Institute of Theoretical Physics, \L{}ojasiewicza 11, 30-348 Krak\'{o}w, Poland}
\author{Marek M. Rams}
\email{marek.rams@uj.edu.pl}
\affiliation{Jagiellonian University, Institute of Theoretical Physics, \L{}ojasiewicza 11, 30-348 Krak\'{o}w, Poland}
\author{Michael Zwolak}
\email{mpz@nist.gov}
\affiliation{Biophysical and Biomedical Measurement Group, Microsystems and Nanotechnology Division, Physical Measurement Laboratory, National Institute of Standards and Technology, Gaithersburg, MD, USA}

\maketitle

\vspace{5\baselineskip}
\tableofcontents

\clearpage
\section{Convergence of the Linear Extrapolation Estimator}

\begin{figure}[h] 
\includegraphics[]{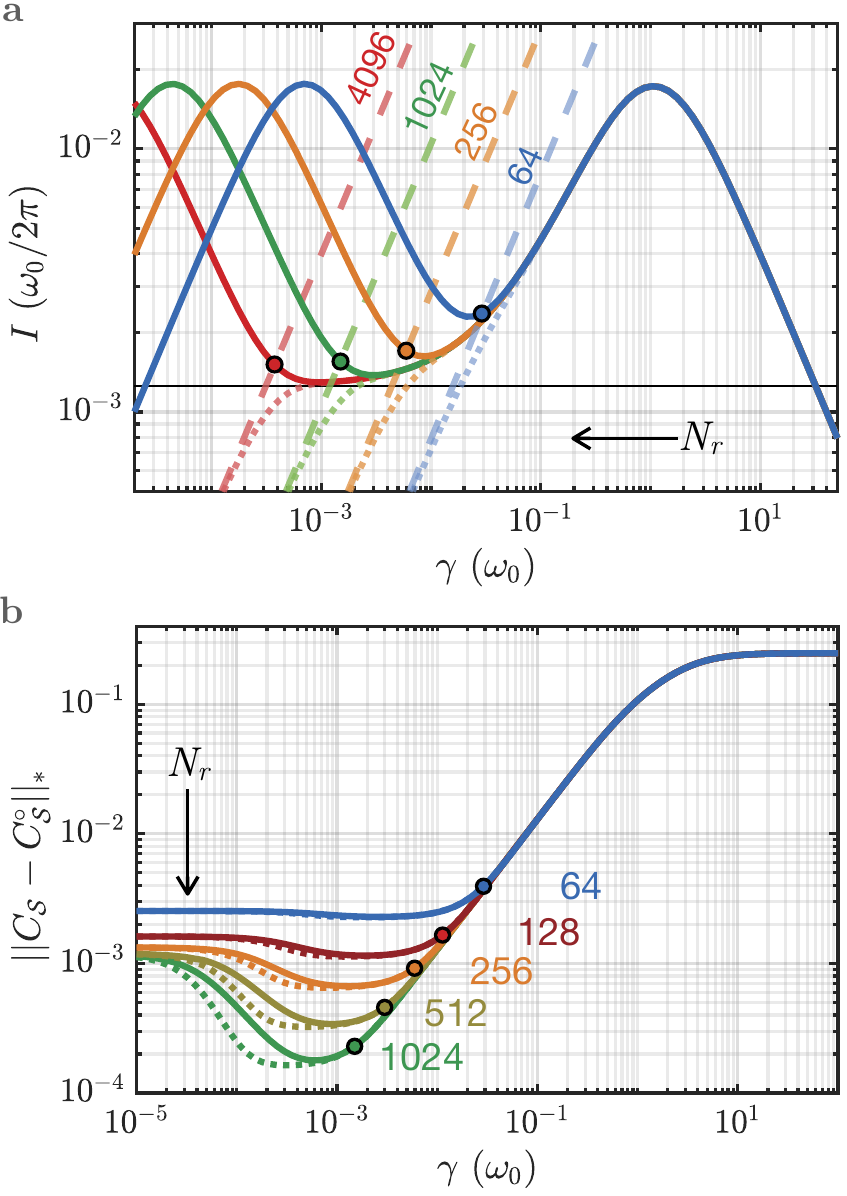}
\caption{{\bf Turnover and the linear extrapolation estimator.} (a) Turnover in the steady--state current $I$ for the non--interacting, dual--site model of Fig.~4 in the primary manuscript.  The on--resonant (solid) and off--resonant (dotted) profiles are shown, along with a linear extrapolation of the off--resonant turnover $\gamma$ (dashed) into the on--resonant turnover.  These profiles intersect at the linear extrapolation estimator $\gamma_\ell$. The relaxation $\gamma_\ell$ settles into the low--to--moderate $\gamma$ shoulder of the virtual anomaly at large $N_r$, leading to the saturation seen in Fig.~4. (b) Turnover of the normalized trace distance between the system correlation matrix $C_\qs$ and its continuum reservoir counterpart $C_\qs^\circ$. The estimator $\gamma_\ell$ is again shown as a circle, which rides a growing minimum in trace distance as $N_r$ is increased.  This implies a scaling profile that does not saturate, as captured by Fig.~4e. The dotted line depicts this trace distance for the real part of $C_\qs$, indicating that current--carrying correlations contribute weakly to the error near $\gamma_\ell$ or $\gamma^\star$. Parameters in (a) and (b) are identical to Fig.~4 in the primary manuscript. }\label{extrap_turnover}
\end{figure}

\clearpage
\section{Current Estimators in Diverse Coupling and Relaxation Regimes}

\par In this extended discussion, we consider several multi--site  models for the system $\qs$ and quantify their transport characteristics in an extended reservoir framework.  In particular, we demonstrate that our intersection ($\gamma_s$) and linear extrapolation ($\gamma_\ell$) estimators for the optimal transport regime $\gamma^\star$ remain applicable under a broad range of parameters.  This underscores their general utility for locating the physical transport regime,  particularly when the turnover profiles have  an ambiguous plateau architecture. 

\subsection{Non--interacting single--site impurity at Weak System--Reservoir Coupling}

\par
\begin{figure}[h] 
\includegraphics[]{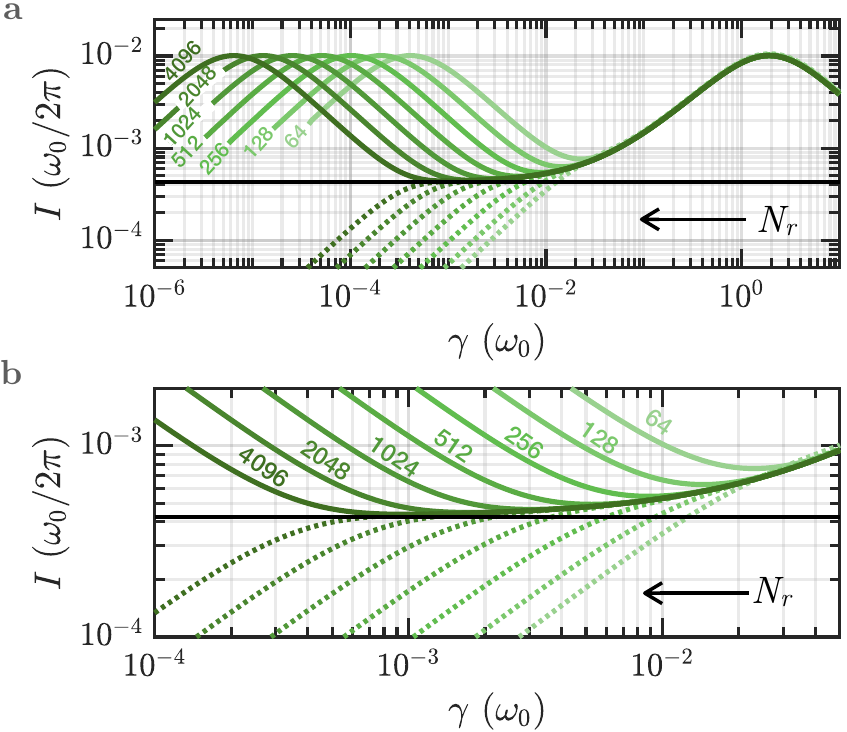}
\caption{{\bf Kramers' turnover for the single--site impurity.} (a) Turnover  in the steady--state current $I$ of the non--interacting single--site impurity model from Eq.~\eqref{eq:singlesite}.  The extended reservoirs have a linear discretization and the relaxation $\gamma_k = \gamma$ is mode independent.  Scaling is assessed with respect to the number $N_r$ of explicit modes in each reservoir for both   on--resonant (solid) and off--resonant (dashed; with a level shift) configurations. (b) Inset of the data from (a) near the physical regime.  The Landauer limit $I^\circ$ for continuum reservoirs is shown by the solid, black horizontal line. Calculations are presented with an on--site frequency $\omega_1 = \omega_0$ and weak system--reservoir coupling $v = \omega_0 / 10$, with a bias of $\mu = \omega_0 / 2$  between $\ql$ and $\qr$ and a temperature $k_B T = \omega_0 / 40$ for reservoir Fermi distributions.  The coupling between system and reservoir sites is provided by the integrated approach (as described in the primary text).}\label{ssm_turnover}
\end{figure}

\begin{figure}[h] 
\includegraphics[]{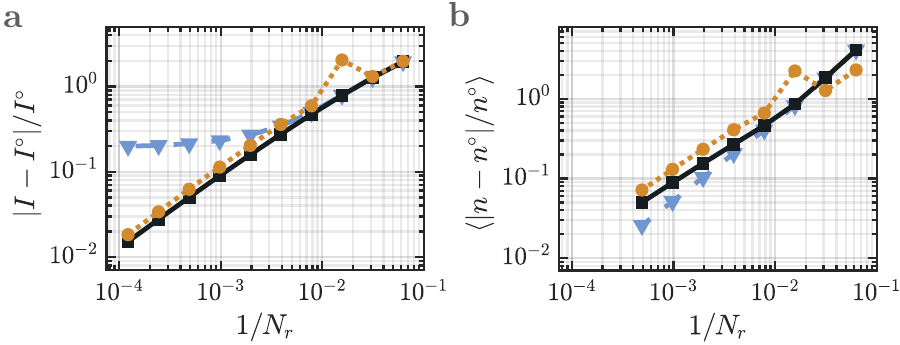}
\caption{{\bf Relaxation estimators for the single--site impurity.}
Convergence of (a) the current error $\vert I - I^\circ\vert / I^\circ$  and (b) on--site density $\vert n - n^\circ\vert / n^\circ$  with respect to $N_r$ for our single--site impurity.  Scaling profiles correspond to  $\gamma_k = \gamma^\star$  (black, square), the  linear extrapolation estimator $\gamma_\ell$ (blue, triangle), and the intersection of shifted/unshifted turnover profiles $\gamma_s$ (tan, circle).  Model parameters are identical to Fig.~\ref{ssm_turnover}, with a linear discretization and integrated system--reservoir couplings.   Error is measured with respect to currents $I^\circ$ and on--site densities  $n^\circ_i$ in the continuum reservoir limit.  }\label{ssm_estimator}
\end{figure}

\begin{figure}[h]
\includegraphics[]{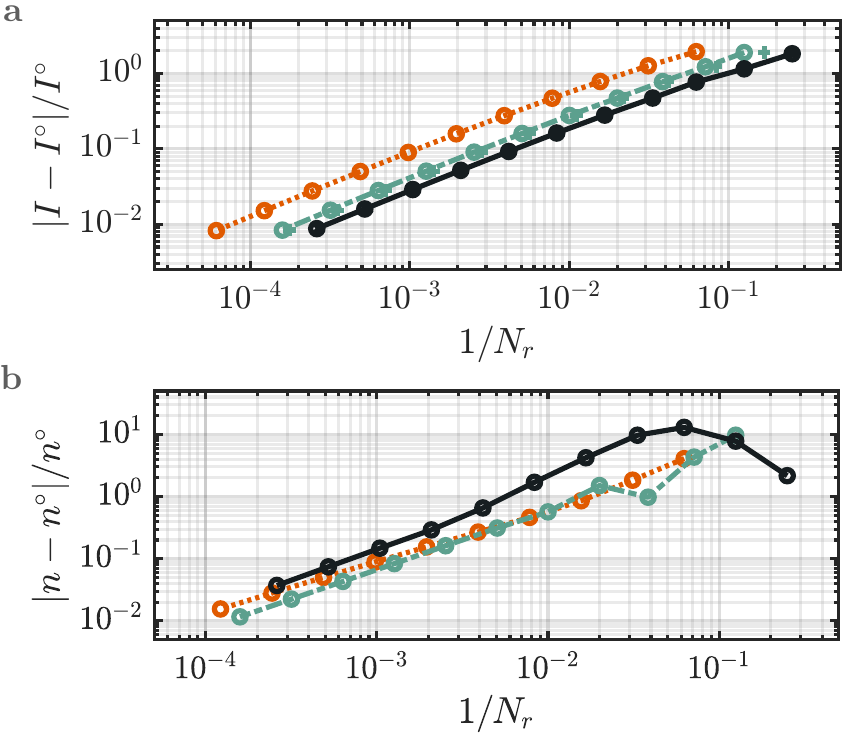}
\caption{{\bf Error and discretization for the single--site impurity.} 
 Convergence of reservoir discretizations for our single--site impurity (Fig.~\ref{ssm_turnover}) when  increasing the number $N_r$ of explicit reservoir sites.  This behavior is quantified through (a) relative error  in the steady--state current $I$ and (b) relative error in the on--site density $n_i$ within $\qs$.  These observables are evaluated at $\gamma^\star$, as defined in the primary manuscript, and the reference current $I^\circ$ is the Landauer limit for continuum reservoirs. Discretizations correspond to the standard linear (orange, dotted line), the linear--logarithmic (green, dashed line), and the linear--inverse (black, solid line) arrangements.  Results are also provided for additional linear--logarithmic and linear--inverse discretizations which are the transform of a 1--d spatial lattice to the energy basis (green and black crosses).   Model parameters are identical to those of Fig.~\ref{ssm_turnover}.} \label{ssm_discretization}
\end{figure}

  The single--site impurity corresponds to a canonical quantum impurity problem (Fig.~1b).  Here, a single mode in $\qs$ is proportionally coupled to left ($\ql$) and right ($\qr$) extended reservoirs with strength $v$.  The system Hamiltonian takes a simple form in terms of $\omega_1$, the on--site system frequency:
\begin{equation}\label{eq:singlesite}
H_\qs =  \hbar \omega_1 \, \cid{1} \ci{1}.
\end{equation}
In this case, the $\gamma$--dependent turnover mimics the dual--site model from the primary manuscript (Fig.~\ref{ssm_turnover}).  Using the linear discretization as a reference, we find an increasingly wide physical `domain of confidence' between virtual and Markovian anomalies as the number $N_r$ of explicit reservoir modes  is increased.  The virtual anomaly also vanishes when modes in $\ql$ and $\qr$ are taken out of resonance, making our $\gamma_\ell$ and $\gamma_s$ estimators applicable.  Taking advantage of this, we find that the intersection estimator $\gamma_s$  is a good predictor for the optimal transport regime $\gamma^\star$ (Fig.~\ref{ssm_estimator}).  We can also quantify convergence of the current at $\gamma^\star$ using other discretizations.  In doing so, we find that the discretizations perform similarly for the one--site (Fig.~\ref{ssm_discretization}) and two--site impuritys (Fig.~\ref{dsm_estimator}) in the weak--coupling limit.  This  agreement  does not extend to on--site densities, where there are discrepancies at small $N_r$ (the linear--logarithmic discretization  performs poorly in this regime). Nonetheless, the discretizations remain comparable when $N_r$ is large (greater than $N_r \approx 128$ sites).

\clearpage 
\subsection{Estimators for the Non--interacting Two--Site Impurity at Weak System--Reservoir Coupling}

\par
\begin{figure}[h] 
\includegraphics[]{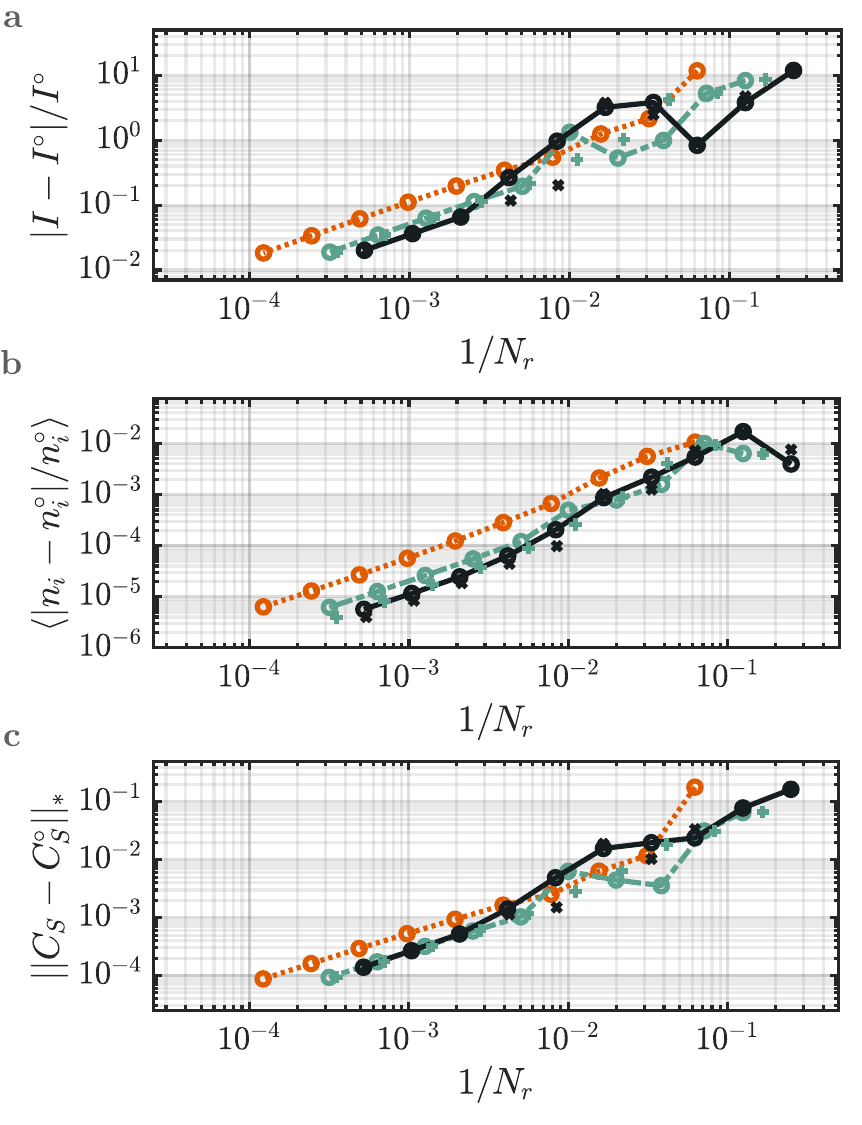}
\caption{{\bf Validation of the intersection estimator across discretizations.} 
The intersection estimator $\gamma_s$ is used to approximate the optimal relaxation regime $\gamma^\star$ in an analysis parallel to Fig.~6 of the primary manuscript.   Convergence is shown for our weakly--coupled ($v = \omega_0 / 10$) two--site impurity (Fig.~\ref{ssm_turnover}) when  increasing the number $N_r$ of explicit reservoir sites.  This behavior is quantified through (a) relative error  in the steady--state current $I$; (b) relative error in the on--site density $n_i$ within $\qs$; and (c) trace distance of the system correlation matrix $C_\qs$ and its continuum reservoir counterpart $C_\qs^\circ$.  Discretizations correspond to the standard linear (orange, dotted line), the linear--logarithmic (green, dashed line), and the linear--inverse (black, solid line) arrangements. Results are also provided for additional linear--logarithmic and linear--inverse discretizations which are the transform of a 1--d spatial lattice to the energy basis (green and black crosses).  Parameters, methodology, and labels are otherwise identical to those of Fig.~6.}\label{dsm_estimator}
\end{figure}

 We  previously quantified the convergence of  reservoir discretizations for a non--interacting two--site impurity at weak--coupling (Fig.~6 of the primary manuscript).   This analysis may be repeated by using $\gamma_s$ to estimate the optimal relaxation $\gamma^\star$.  In doing so, we find that the distinction between discretizations is now even less apparent.  Nonetheless, $\gamma_s$ gives a robust estimate irrespective of context (Fig.~\ref{dsm_estimator}), supporting its use with  many--body impurities.

\clearpage
\subsection{Non--interacting two--site impurity at Strong System--Reservoir Coupling}

\par
\begin{figure}[h] 
\includegraphics[]{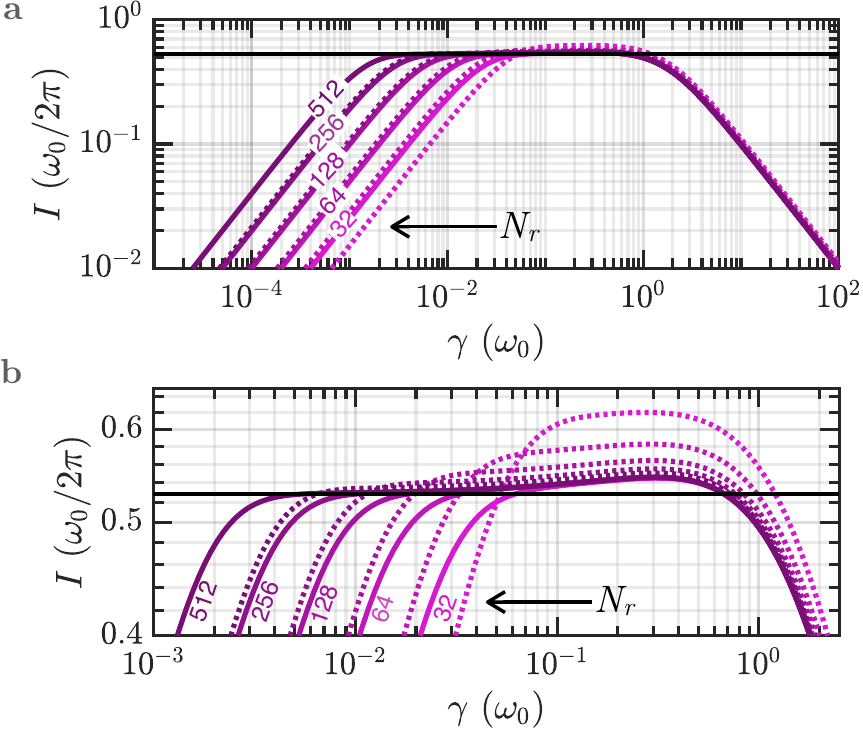}
\caption{{\bf Kramers' turnover for the strongly--coupled two--site impurity.} (a) Turnover  in the steady--state current $I$ of the non--interacting two--site impurity model from Fig 3b of the primary manuscript, but now at strong $\qs-\ql\qr$ coupling.  The extended reservoirs have a linear discretization and the relaxation $\gamma_k = \gamma$ is mode independent.  Scaling is assessed with respect to the number $N_r$ of explicit modes in each reservoir for both   on--resonant (solid) and off--resonant (dashed; with a level shift) configurations. (b) Inset of the data from (a) near the physical regime.  The Landauer limit $I^\circ$ for continuum reservoirs is shown by the solid, black horizontal line. Calculations are presented with on--site frequencies $\omega_{\qs,j} = 0\,\cdot\, \omega_0$, a coupling of $v_{12} = (1 + \sqrt{2}) \omega_0 /4$ between system sites, and strong $\qs$-$\ql\qr$ coupling $v = \omega_0 / 2$.   A bias of $\mu = \omega_0 / 2$ is applied  between $\ql$ and $\qr$ and a temperature $k_B T = \omega_0 / 40$ is used for reservoir Fermi distributions.  Couplings are provided by the integrated approach.} \label{dsm_sc_turnover}
\end{figure}

 Our primary analysis was performed for a non--interacting, two--site impurity at weak  system--reservoir coupling ($v = \omega_0/10$).  It is instructive to analyze the same system when the coupling is strong ($v = \omega_0/2$)  and the turnover  becomes markedly different (Fig.~\ref{dsm_sc_turnover}).   At weak coupling, our single-- and double--site systems  have a well--defined valley between current turnover anomalies.  The bottom of this feature approaches the continuum--limit $I^\circ$ at the optimal relaxation $\gamma^\star$ and broadens as $N_r$ is increased.   The behavior changes at strong coupling, where there is now a flat turnover plateau that intersects $I^\circ$ so that $I(\gamma^\star) = I^\circ$ exactly.  This regime is difficult to identify by visual inspection due to a lack of prominent plateau features (Fig.~\ref{dsm_sc_turnover}b), though it does lie at the small--$\gamma$ side due to a weak virtual anomaly.

\par
\begin{figure}[h] 
\includegraphics[]{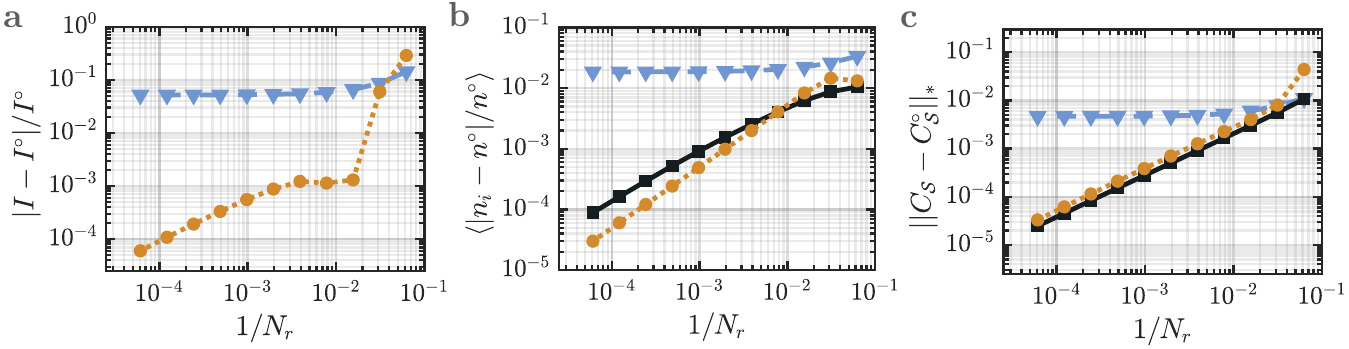}
\caption{{\bf Relaxation estimators for the strongly coupled two--site impurity.}
Convergence of (a) the current error $\vert I - I^\circ\vert / I^\circ$; (b) the mean error of the on--site density $\rangle \vert n_i - n^\circ_i \vert / n^\circ_i \langle$; and (c) the normalized trace distance between the system correlation matrix $C_\qs$ and its continuum reservoir counterpart $C_\qs^\circ$.   Scaling is provided with respect to the number of explicit modes in each reservoir $N_r$ and estimated using $\gamma_k = \gamma^\star$  (black, square), the  linear extrapolation estimator $\gamma_\ell$ (blue, triangle), and the intersection of shifted/unshifted turnover profiles $\gamma_s$ (tan, circle).  Model parameters are identical to Fig.~\ref{dsm_sc_turnover}, with a linear discretization and integrated system--reservoir couplings.   Error is measured with respect to currents $I^\circ$ and on--site densities  $n^\circ_i$ in the Landauer limit of continuum reservoirs.  The $I(\gamma^\star)$ scaling is not provided for (a) since the plateau intersects the Landauer limit and thus the result is inherently exact.  }\label{dsm_sc_estimator}
\end{figure}

 Despite the change in plateau architecture, the intersection estimator $\gamma_s$ remains robust and applicable (Fig.~\ref{dsm_sc_estimator}).  In fact, this estimator affords currents $I(\gamma_s)$ that steadily approach the continuum reservoir limit  $I^\circ$ as $N_r$ is increased. The linear extrapolation  estimator $\gamma_\ell$ consistently underestimates the current, intersecting the turnover  near the mid--plateau upturn (as we move toward smaller $\gamma$) that is associated with the virtual anomaly. Interestingly, this estimator performs poorly for the on--site densities $n_i$, with an error that has already saturated at small $N_r$.  Similar behavior is seen when using  $C_\qs$ as a measure for the full state of the system.  Linear extrapolation from the off--resonant profile now intersects the on--resonant plateau at a point away from the strong downturn regime captured in Fig.~\ref{extrap_turnover}b.

\par
\begin{figure}[h] 
\includegraphics[]{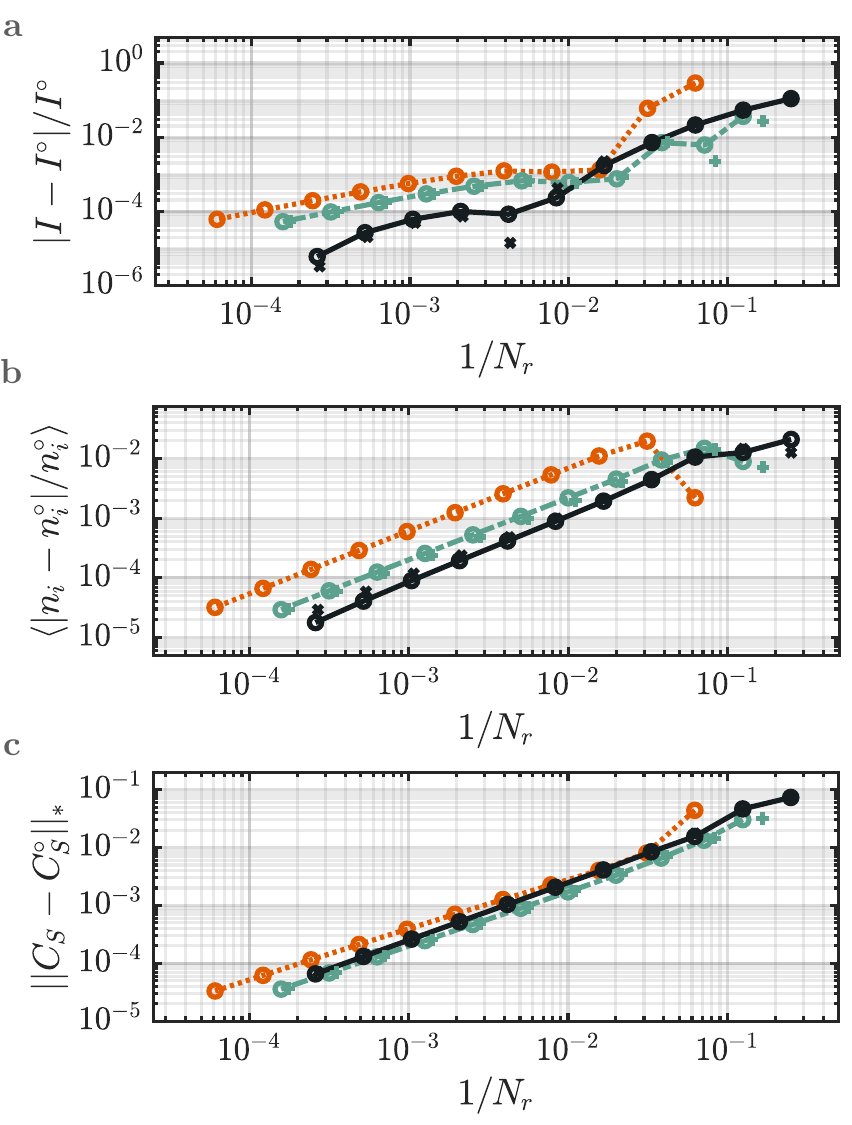}
\caption{{\bf Estimator validation across discretizations for the strongly coupled two--site impurity.} 
The intersection estimator $\gamma_s$ is used to approximate the optimal relaxation regime $\gamma^\star$ in an analysis that parallels Fig.~6 of the primary manuscript.   Convergence is shown for our strongly--coupled ($v = \omega_0 / 2$) two--site impurity (Fig.~\ref{ssm_turnover}) when  increasing the number $N_r$ of explicit reservoir sites.  This behavior is quantified through (a) relative error  in the steady--state current $I$; (b) relative error in the on--site density $n_i$ within $\qs$; and (c) the trace between the system correlation matrix $C_\qs$ and its continuum reservoir counterpart $C_\qs^\circ$. The reference values ($I^\circ$, $n_i^\circ$ and $C_\qs^\circ$) are taken in the continuum reservoir limit. Discretizations correspond to the standard linear (orange, dotted line), the linear--logarithmic (green, dashed line), and the linear--inverse (black, solid line) arrangements. Results are also provided for additional linear--logarithmic and linear--inverse discretizations which are the transform of a 1--d spatial lattice to the energy basis (green and black crosses).  Parameters, methodology, and labels are otherwise identical to those of Fig.~6.}
\label{dsm_sc_discretization}
\end{figure}

\begin{figure}[h] 
\includegraphics[]{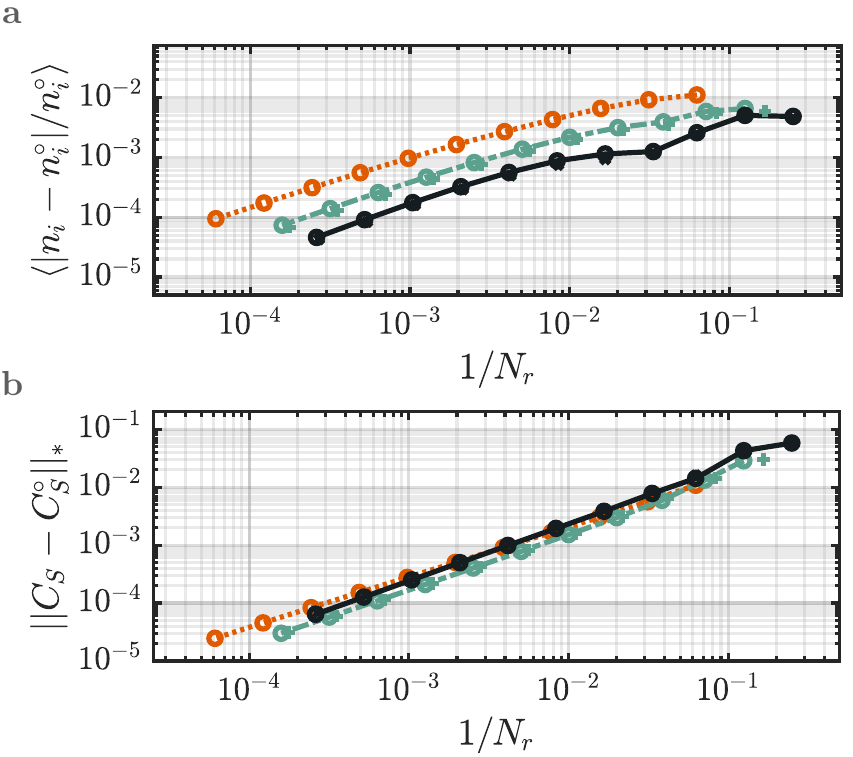}

\caption{{\bf Error and discretization for the strongly coupled two--site impurity.} Convergence of reservoir discretizations for our two--site impurity (Fig.~\ref{ssm_turnover}) when  increasing the number $N_r$ of explicit reservoir sites at strong $\qs$-$\ql\qr$  coupling ($v = \omega_0/2$).  These observables are evaluated at $\gamma^\star$, as defined in the primary manuscript. This behavior is quantified through  (a) relative error in the on--site density $n_i$ within $\qs$ and (b) the normalized trace distance between the system correlation matrix $C_\qs$ and its continuum reservoir counterpart $C_\qs^\circ$. The reference values ($n_i^\circ$ and $C_\qs^\circ$) are taken in the continuum reservoir limit. Discretizations correspond to the standard linear (orange, dotted line), the linear--logarithmic (green, dashed line), and the linear--inverse (black, solid line) arrangements. Results are also provided for additional linear--logarithmic and linear--inverse discretizations which are the transform of a 1--d spatial lattice to the energy basis (green and black crosses).  Model parameters are identical to those of Fig.~6.}\label{dsm_sc_discretization2}
\end{figure}

 The estimator $\gamma_s$ also delivers robust performance for different discretizations at strong coupling, as shown in Fig.~\ref{dsm_sc_discretization}. That is, all mode arrangements accurately assess key physical quantities. Although currents are formally exact for the optimal estimator $\gamma^\star$  in this strongly--coupled limit, we can nonetheless compare to mean errors of the densities and correlation matrices calculated at $\gamma^\star$.  In doing so, we observe favorable performance in $\gamma_s$ when estimating $\gamma^\star$ in terms of both error magnitude and the suitability of different discretizations (Fig.~\ref{dsm_sc_discretization2}).  

\clearpage

\clearpage

\clearpage

\subsection{ Three--Site Model}

\par
\begin{figure}[h]
\includegraphics[]{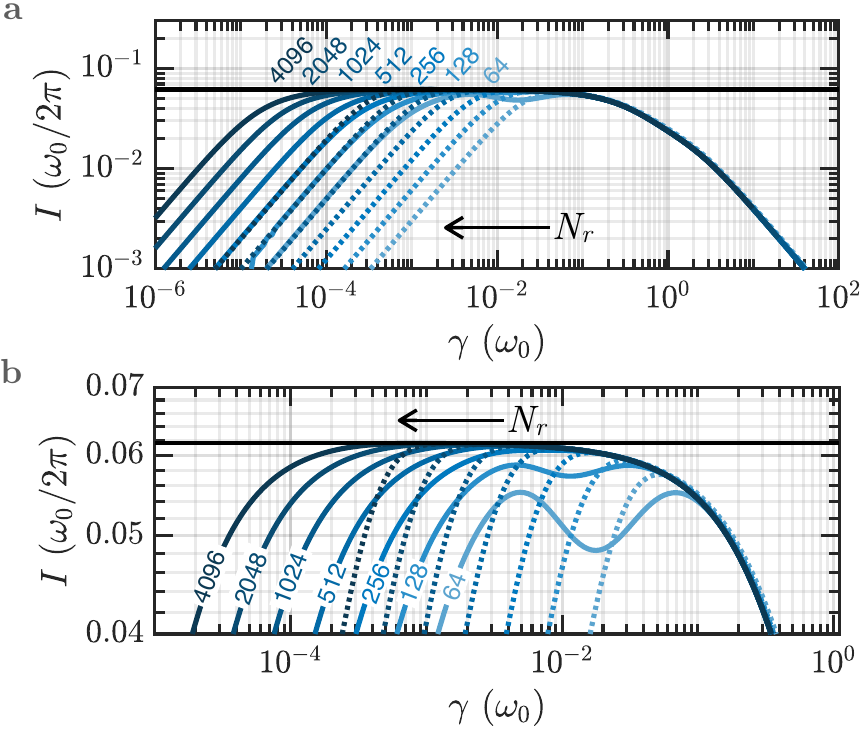}
\caption{{\bf Kramers' turnover for a weakly coupled three--site impurity.} (a) Turnover  in the steady--state current $I$ for the non--interacting three--site  model of Eq.~\eqref{eq:threesite}, taken  at weak $\qs$-$\ql\qr$ coupling ($v =  \omega_0/10$).  The extended reservoirs have a linear discretization and the relaxation $\gamma_k = \gamma$ is mode independent.  Scaling is assessed with respect to the number $N_r$ of explicit modes in each reservoir, for both on--resonant (solid) and off--resonant (dashed; with a level shift) configurations. (b) Inset of the data from (a) near the physical regime.  The Landauer limit $I^\circ$ for continuum reservoirs is shown by the solid, black horizontal line. Calculations are presented with  on--site  system frequencies $\omega_{j} = 0 \,\cdot\, \omega_0$, couplings of $v_{{12}} = v_{{23}} = (1 + \sqrt{2})\omega_0 /4$ between system sites, and strong system--reservoir coupling $v = \omega_0 / 2$.   A bias of $\mu = \omega_0 / 2$ is applied  between $\ql$ and $\qr$ and a temperature $k_B T = \omega_0 / 40$ is used for reservoir Fermi distributions.  Couplings are provided by our integrated approach.}\label{tsm_turnover}
\end{figure}

\begin{figure}[h] 
\includegraphics[]{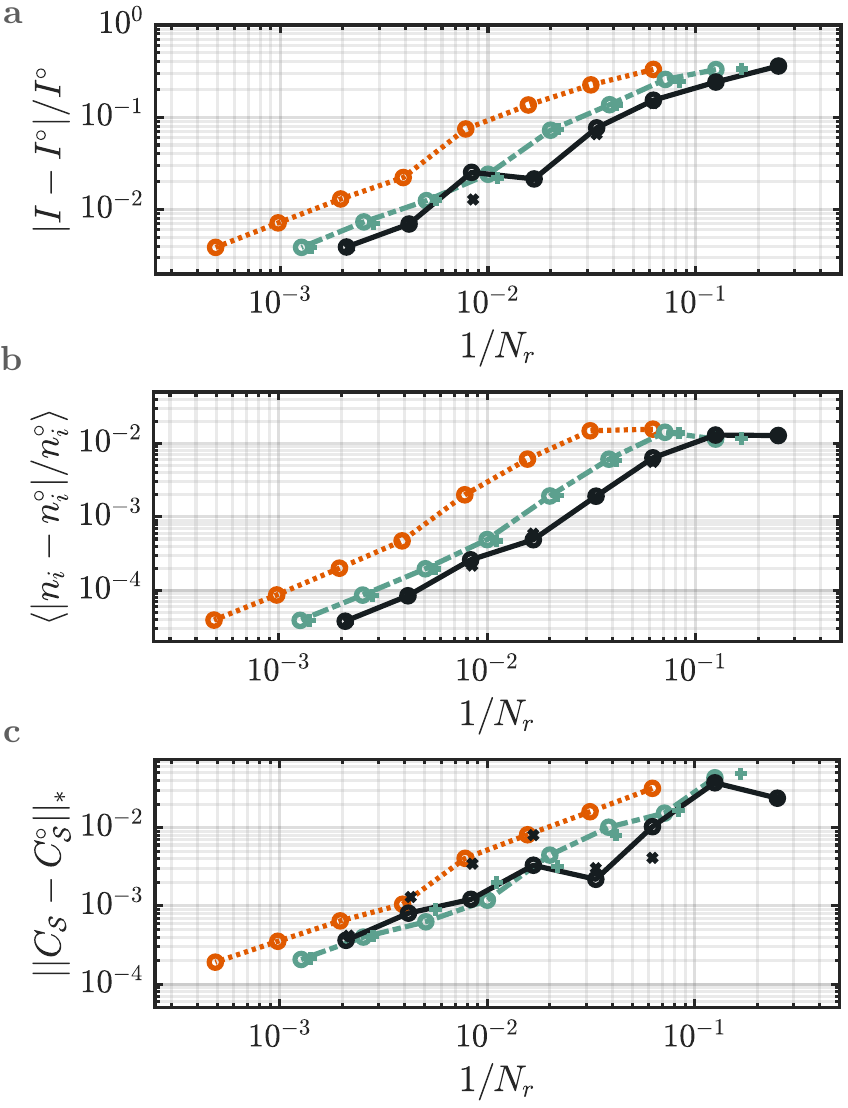}
\caption{{\bf Estimator validation across discretizations for the weakly coupled three--site impurity.} 
The intersection estimator $\gamma_s$ is used to approximate the optimal relaxation regime $\gamma^\star$ in an analysis parallel to Fig.~6 of the primary manuscript.   Convergence is shown for a weakly coupled ($v = \omega_0 / 10$) three--site impurity (Eq.~\eqref{eq:threesite} and Fig.~\ref{tsm_turnover}) when  increasing the number $N_r$ of explicit reservoir modes.  This behavior is quantified through (a) relative error  in the steady--state current $I$; (b) relative error in the on--site density $n_i$ within $\qs$; and (c) the trace between the system correlation matrix $C_\qs$ and its continuum reservoir counterpart $C_\qs^\circ$. The reference values ($I^\circ$, $n_i^\circ$ and $C_\qs^\circ$) are taken in the continuum reservoir limit.  Discretizations correspond to the standard linear (orange, dotted line), the linear--logarithmic (green, dashed line), and the linear--inverse (black, solid line) arrangements. Results are also provided for additional linear--logarithmic and linear--inverse discretizations which are the transform of a 1--d spatial lattice to the energy basis (green and black crosses).  Parameters, methodology, and labels are otherwise identical to those of Fig.~6.}
 \label{tsm_discretization}
\end{figure}

 The three--site impurity is a more complex case, corresponding to a linear chain of sequentially--coupled sites. These connect to the semi--infinite reservoir $\ql$ at the first site of the chain and to the reservoir $\qr$ at the last site. The Hamiltonian for this arrangement reads

\begin{equation}\label{eq:threesite}
H_\qs = \sum_{j=1}^3 \hbar \omega_j\, \cid{j}\ci{j} + \hbar v_{{12}} \, (\cid{1} \ci{2} + \text{h.c.}) + \hbar v_{{23}} \, (\cid{2} \ci{3} + \text{h.c.})
\end{equation}

\noindent which, in essence, is a three--site tight--binding model. For simplicity, we adopt on--site energies of $\omega_j = 0 \cdot \omega_0$, system--site couplings $ v_{{12}} =  v_{{23}} = (1 + \sqrt{2})\omega_0 /4$ and a weak system--reservoir coupling scale $v = \omega_0 / 10$.  Unlike the other cases studied, this model has a current turnover profile that approaches the continuum limit $I^\circ$ from below (Fig.~\ref{tsm_turnover}).  Such behavior precludes estimators for $\gamma^\star$ that  minimize $\vert I(\gamma) - I^\circ \vert / I^\circ$, since the current maxima of the anomalies (both Markovian and virtual) approach $I^\circ$ more rapidly than the interstitial, physical regime.  We nonetheless find that the intersection estimator $\gamma_s$ remains effective for identifying the physical regime of the plateau, exhibiting robust scaling for all quantities with respect to increasing $N_r$ (Fig.~\ref{tsm_discretization}).  Within this context, the linear--inverse discretization continues to exhibit robust performance for all observables.  Taken together, our observations underscore how our intersection estimator $\gamma_s$ is generally robust when identifying optimal transport regimes --- a fact that holds irrespective of plateau architecture or the manner in which the thermodynamic limit is approached.  

\clearpage

\clearpage

\section{Alternative Linear--Logarithmic Discretizations}

\subsection{Method I}

\par The logarithmic discretization from Eq.~(3) and Fig.~6 is based on an influence measure that maintains continuity across the bias window edge.  While convenient, continuity is not required.  In fact, we can introduce a more general influence measure, 

\begin{equation}
\zi_\text{log-alt1}(\omega) = \theta\left(\frac{\mu}{2} - \vert \omega \vert\right) + \frac{\alpha}{\vert\omega\vert} \theta\left(\vert \omega \vert - \frac{\mu}{2}\right), \\
\end{equation}

\noindent that is specified by an independent discretization scale $\Lambda = \omega_{k+1} / \omega_k$.  This is imposed by extending  the bias window level spacing $\Delta_0$ to the first logarithmic mode (e.g., $\Lambda = 1 + 2\Delta_0 / \mu$).  We then invoke the definition of our influence measure,
\begin{equation}\label{eq:linLogAlt1}
\int_{\mu/2}^{\mu/2 + \Delta_0} \zi_\text{log-alt1}(\omega) \, d\omega = \Delta_0 .
\end{equation}

\noindent to establish a relationship $\alpha = \frac{\Delta_0}{\log \Lambda}$ between these quantities. Unlike the approach of Eq.~(3), this approach has two independent parameters and, thus, while it retains a linear and a logarithmic sector, it allows for them to be adjusted separately.

\par
\begin{figure}[t!]  
\begin{center}
\includegraphics[scale=0.95]{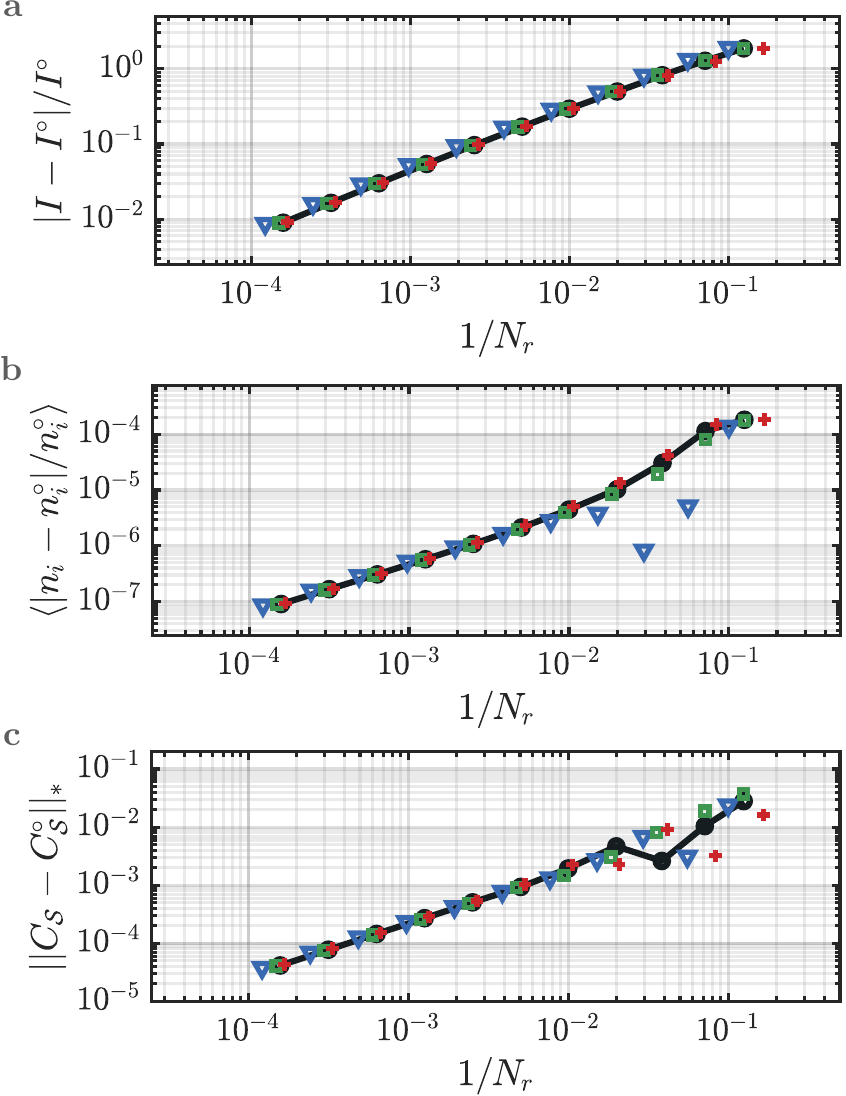}
\end{center}
\caption{{\bf Error and discretization for an alternate linear--logarithmic discretization.} Convergence of the alternate linear--logarithmic discretization of Eq.~\eqref{eq:linLogAlt1} at different discretization scales $\Lambda$.  This behavior is quantified through (a) relative error in the steady--state current $I[\gamma_k(\alpha^\star)]$; (b) relative error in the mean on--site density $n_i$ within $\qs$; and (c) the normalized trace distance between correlation matrix $C_\qs$ for $\qs$ and its infinite reservoir counterpart $C_\qs^\circ$. Discretizations correspond to the standard  linear--logarithmic influence measure of Eq.~(3) of the primary manuscript (black, solid line) and the alternate influence measure from Eq.~\eqref{eq:linLogAlt1} with $\Lambda = 2.0$ (blue, triangles), $\Lambda = 2.5$ (green, squares), and $\Lambda = 3.0$ (red, crosses).   All data are from the non--interacting, two--site Hamiltonian of Fig.~3b at weak--coupling ($v_0 = \omega_0 / 10$), with integrated system--reservoir couplings, and relaxations $\gamma = \alpha \langle \Delta_k\rangle_{\qb}$ determined by the mean mode spacing within the bias window $\mu = \omega_0 / 2$.}
\label{fig:scalingAltLinLog}
\end{figure}

 We find this alternate arrangement is comparable to the standard linear--logarithmic discretization for practical transport calculations (Fig.~\ref{fig:scalingAltLinLog}).  When working at a discretization scale ($\Lambda = 2.0$ to $\Lambda = 3.0$) that is typical for the numerical renormalization group (NRG), we find only subtle variations in the calculated currents.  While some  deviations are seen at small $N_r$ for  on--site densities and the system state, these vanish for a large number of reservoir sites.

\clearpage
\subsection{Method II}

\par Here, we assess a third logarithmic discretization that has been used in prior many--body transport calculations (see Ref.~\cite{schwarz_nonequilibrium_2018}).  We construct this by subdividing the reservoir bandwidth $\mathcal{W}$ into bins of width $\Delta_k = \vert \Omega_{k+1} - \Omega_k \vert$, where the frequencies $\Omega_k$ are indexed by $k\in \mathbb{Z}$.  The sign of $k$ corresponds to the sign of the associated frequency.   We then arrange the $\Omega_k$ to give $N_\mu = \mu/\Delta_0$ linearly spaced bins inside the bias window.  Modes far outside the bias window  will be distributed logarithmically  (e.g., $\Omega_{k+1} / \Omega_k = \Lambda$ for $k > 0$) with $\Lambda$ setting the logarithmic discretization scale. As with Method I, this permits independent adjustment of the linear and logarithmic sectors.   The full distribution of $\Omega_k$ is then defined to interpolate between these limits,

\begin{equation}\label{eq:linLogVD1}
\Omega_k =\begin{cases}
\Delta_0 \cdot k, & \text{if } \vert k \vert \leq N_\mu / 2, \\
 \Delta_0 \cdot \left(\frac{\sinh \left[\left(k \mp \frac{N_\mu}{2}\right) \log \Lambda\right]}{\log \Lambda} \pm \frac{N_\mu}{2} \right) & \text{if } \vert k\vert > N_\mu / 2.
\end{cases}
\end{equation}

\noindent  This set is used to specify mode placement, assuming that the extremal bins bounded by the band edges at $\pm \mathcal{W} / 2$.  We take two approaches to define the modes $\omega_k$.  In the first, we place modes at the midpoint of each bin so that $\omega_k = [\Omega_{k+1} + \Omega_k]/2$.  The second follows Ref.~\cite{schwarz_nonequilibrium_2018}, where we set

\begin{equation}\label{eq:linLogVD2}
\omega_k =\begin{cases}
\frac{\Omega_{k+1} - \Omega_k}{\log (\Omega_{k+1} / \Omega_k)} & \text{if } |\Omega_k\vert, \vert\Omega_{k+1}\vert > \mu/2,\\
\frac{1}{2}[\Omega_{k+1} + \Omega_{k}] & \text{otherwise}.
\end{cases}
\end{equation} 

\noindent We introduce  couplings $v_{jk}$ that reproduce the integrated weight within each bin and relaxations $\gamma_k$ that are specified by the mode spacing in the bias window (as used in Fig.~6).

\par
\begin{figure}[t!]  
\begin{center}
\includegraphics[scale=0.95]{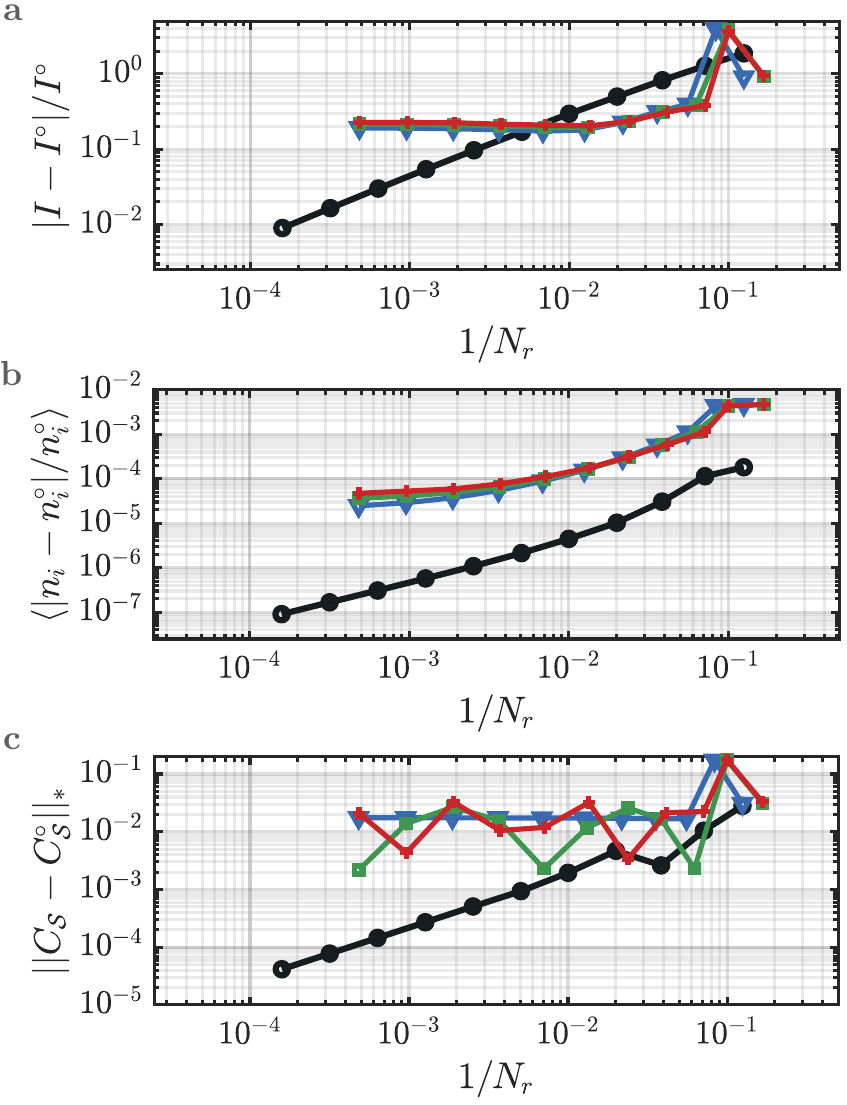}
\end{center}
\caption{{\bf Error and discretization for an alternate (Method I) linear--logarithmic discretization.} Convergence of the alternate linear--logarithmic discretization of Eq.~\eqref{eq:linLogVD1} at different discretization scales $\Lambda$.  Here, the reservoir modes are defined by placing them at the center of each discretization bin. This behavior is quantified through (a) relative error in the steady--state current $I[\gamma_k(\alpha^\star)]$; (b) relative error in the mean on--site density $n_i$ within $\qs$; and (c) the normalized trace distance between correlation matrix $C_\qs$ for $\qs$ and its infinite reservoir counterpart $C_\qs^\circ$. Discretizations correspond to the standard  linear--logarithmic influence measure of Eq.~(3) of the primary manuscript (black, circles) and the alternate influence measure from Eq.~\eqref{eq:linLogVD1} with $\Lambda = 2.0$ (blue, triangles), $\Lambda = 2.5$ (green, squares), and $\Lambda = 3.0$ (red, crosses).   All data are from the non--interacting, two--site Hamiltonian of Fig.~3b at weak--coupling ($v_0 = \omega_0 / 10$), with integrated system--reservoir couplings, and relaxations $\gamma = \alpha \langle \Delta_k\rangle_{\qb}$ determined by the mean mode spacing within the bias window $\mu = \omega_0 / 2$.}
\label{fig:scalingLinLogVD1}
\end{figure}

\begin{figure}[t!]  
\begin{center}
\includegraphics[scale=0.95]{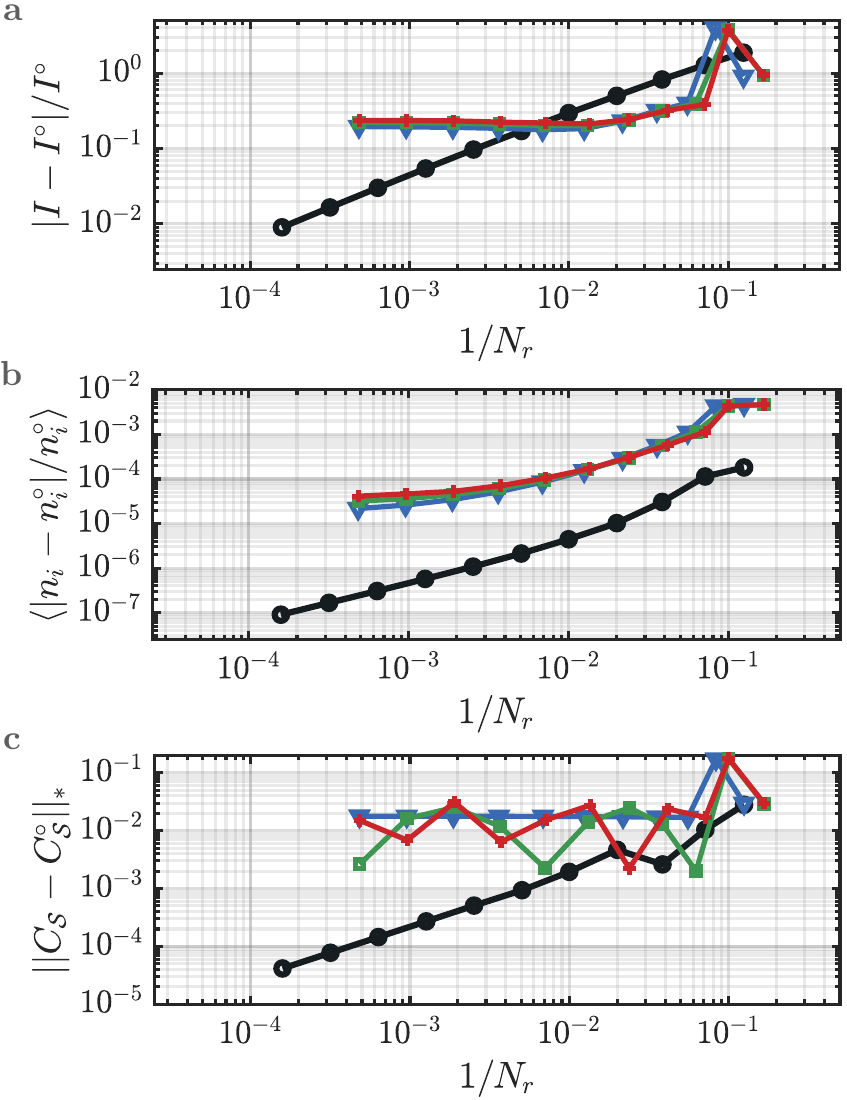}
\end{center}
\caption{{\bf Error and discretization for an alternate (Method II) linear--logarithmic discretization.} Convergence of the alternate linear--logarithmic discretization of Eq.~\eqref{eq:linLogVD1} at different discretization scales $\Lambda$.  Here, the reservoir modes are defined according to Eq.~\eqref{eq:linLogVD2}. This behavior is quantified through (a) relative error in the steady--state current $I[\gamma_k(\alpha^\star)]$; (b) relative error in the mean on--site density $n_i$ within $\qs$; and (c) the normalized trace distance between correlation matrix $C_\qs$ for $\qs$ and its infinite reservoir counterpart $C_\qs^\circ$. Discretizations correspond to the standard  linear--logarithmic influence measure of Eq.~(3) of the primary manuscript (black, circles) and the alternate influence measure from Eq.~\eqref{eq:linLogVD1} with $\Lambda = 2.0$ (blue, triangles), $\Lambda = 2.5$ (green, squares), and $\Lambda = 3.0$ (red, crosses).   All data are from the non--interacting, two--site Hamiltonian of Fig.~3b at weak--coupling ($v_0 = \omega_0 / 10$), with integrated system--reservoir couplings, and relaxations $\gamma = \alpha \langle \Delta_k\rangle_{\qb}$ determined by the mean mode spacing within the bias window $\mu = \omega_0 / 2$.}
\label{fig:scalingLinLogVD2}
\end{figure}

 In practice, the discretization specified by Eq.~\eqref{eq:linLogVD1} is insensitive  to how modes are placed within a given frequency bin (Figs.~\ref{fig:scalingLinLogVD1} and \ref{fig:scalingLinLogVD2}, respectively). Furthermore, it scales more efficiently for currents than the standard linear--logarithmic method from Eq.~(3), provided that  $N_r$ is small.  This is not surprising, as transport is dominated by states in the bias window and the method of Eq.~\eqref{eq:linLogVD1} favors this region.  Nonetheless, the observed benefit saturates beyond the number of modes typical for most many--body simulations  ($N_r \approx 128$), while the saturation point has a moderate dependence on the discretization scale $\Lambda$.  Such behavior is also expected, as the bandwidth outside the bias window is heavily coarse--grained and relevant correlations may be missed.  This is also reflected through a saturation in on--site densities and without a well-defined trend for the error in the system state.  

This behavior is not unexpected for a two parameter family. The lin-log discretization of the main text links $\Delta_0$ and $\Lambda$, taking the continuum limit of both together. Method I and II, however, have independent adjustment of these two parameters. When they are not balanced, this can cause a saturation in error as further decrease of one parameter (towards the continuum limit) can not capture what is missing due to the fixed value of the other parameter.

\clearpage

\clearpage

